\documentclass[aps,twocolumn,superscriptaddress]{revtex4-2}
\usepackage{braket}
\usepackage{dcolumn}
\usepackage{graphicx} 

\usepackage[normalem]{ulem}

\usepackage{placeins}
\usepackage{listings}

\usepackage{physics} %

\usepackage{braket} 

\usepackage{amsmath,amssymb}
\usepackage{amsthm}
\usepackage{bm}
\usepackage{graphicx}
\usepackage{ascmac}

\usepackage{mathrsfs}
\usepackage{color}
\usepackage{booktabs}
\usepackage{multirow}
\usepackage{orcidlink}
\usepackage{todonotes}

\usepackage{mathtools}
\usepackage{textcomp}

\usepackage{natbib}
\usepackage{tikz}
\usetikzlibrary{decorations.pathreplacing}

\newif\ifshowmodelgrid
 \showmodelgridfalse   

\newcommand{\modelgrid}{%
  \ifshowmodelgrid
    \begin{pgfinterruptboundingbox}
      \def\gridxmin{-7.5}
      \def\gridxmax{8.5}
      \def\gridymin{-3.5}
      \def\gridymax{3.5}

      \def\gridlabelx{-8.0}
      \def\gridlabely{-3.35}

      \draw[help lines, step=0.5, gray!55, very thin, opacity=0.35]
        (\gridxmin,\gridymin) grid (\gridxmax,\gridymax);

      \draw[help lines, step=1, gray!75, thin, opacity=0.55]
        (\gridxmin,\gridymin) grid (\gridxmax,\gridymax);

      \foreach \x in {-7,-6,...,8}
        \node[
          font=\scriptsize,
          anchor=north,
          fill=white,
          fill opacity=0.85,
          text opacity=1,
          inner sep=1pt
        ] at (\x,\gridlabely) {\x};

      \foreach \y in {-3,-2,...,3}
        \node[
          font=\scriptsize,
          anchor=west,
          fill=white,
          fill opacity=0.85,
          text opacity=1,
          inner sep=1pt
        ] at (\gridlabelx,\y) {\y};
    \end{pgfinterruptboundingbox}
  \fi
}

\newcommand{\ideal}{\mathrm{ideal}}

\newcommand{\tgt}{\mathrm{t}}
\newcommand{\src}{\mathrm{s}}
\newcommand{\dom}{\mathrm{dom}}

\newcommand{\RR}{\mathbb{R}}

\newcommand{\cE}{\mathcal{E}}

\newcommand{\cS}{\mathcal{S}}

\newcommand{\cY}{\mathcal{Y}}
\newcommand{\cL}{\mathcal{L}}
\newcommand{\cM}{\mathcal{M}}

\newcommand{\modelrule}{%
  \specialrule{\lightrulewidth}{0pt}{0pt}
  \specialrule{\lightrulewidth}{1pt}{0pt}
  \addlinespace
}

\definecolor{grey}{rgb}{.5,.5,.5}

\usepackage{ulem}

\DeclareMathOperator*{\argmax}{argmax}

\usepackage{enumitem}

\begin{document}
\title{Learning from imperfect quantum data via unsupervised domain adaptation \\ with classical shadows}
\author{Kosuke Ito}
\email[]{kosuke.ito@quantum.keio.ac.jp}
\affiliation{Advanced Material Engineering Division, Toyota Motor Corporation, 1200 Mishuku, Susono, Shizuoka 410-1193, Japan}
\affiliation{Quantum Computing Center, Keio University, Hiyoshi 3-14-1, Kohoku, Yokohama, Kanagawa 223-8522, Japan}

\author{Akira Tanji}
\affiliation{Department of Applied Physics and Physico-Informatics, Keio University, Hiyoshi 3-14-1, Kohoku, Yokohama, Kanagawa 223-8522, Japan}

\author{Hiroshi Yano}
\affiliation{Quantum Computing Center, Keio University, Hiyoshi 3-14-1, Kohoku, Yokohama, Kanagawa 223-8522, Japan}

\author{Yudai Suzuki}
\affiliation{Institute of Physics, \'{E}cole Polytechnique F\'{e}d\'{e}rale de
Lausanne (EPFL), Lausanne, Switzerland}
\affiliation{Centre for Quantum Science and Engineering, \'{E}cole Polytechnique F\'{e}d\'{e}rale de Lausanne (EPFL), Lausanne, Switzerland}
\affiliation{Quantum Computing Center, Keio University, Hiyoshi 3-14-1, Kohoku, Yokohama, Kanagawa 223-8522, Japan}

\author{Naoki Yamamoto}
\affiliation{Quantum Computing Center, Keio University, Hiyoshi 3-14-1, Kohoku, Yokohama, Kanagawa 223-8522, Japan}
\affiliation{Department of Applied Physics and Physico-Informatics, Keio University, Hiyoshi 3-14-1, Kohoku, Yokohama, Kanagawa 223-8522, Japan}


\begin{abstract}
    Learning from quantum data using classical machine learning models has emerged as a promising paradigm toward realizing quantum advantages.
    Despite extensive analyses on their performance, clean and fully labeled quantum data from the target domain are often unavailable in practical scenarios, forcing models to be trained on data collected under conditions that differ from those encountered at deployment.
    This mismatch highlights the need for new approaches beyond the common assumptions of prior work.
    In this work, we address this issue by employing an unsupervised domain adaptation framework for learning from imperfect quantum data.
    Specifically, by leveraging classical representations of quantum states obtained via classical shadows, we perform unsupervised domain adaptation entirely within a classical computational pipeline once measurements on the quantum states are executed.
    We numerically evaluate the framework on quantum phases of matter and entanglement classification tasks under realistic domain shifts.
    Across both tasks, our method outperforms source-only non-adaptive baselines and target-only unsupervised learning approaches, demonstrating the practical applicability of domain adaptation to realistic quantum data learning.
\end{abstract}

\maketitle

\section{Introduction}

Learning from quantum data with classical machine learning has emerged as a promising route for extracting useful information from complex quantum systems including many-body quantum systems. 
A key motivation behind this approach is that the access to quantum data changes the computational landscape of prediction tasks, potentially rendering classically-hard tasks tractable.
Recent theoretical work has shown that, for suitable problems, learning with quantum data can be efficient even when the corresponding task without such data is classically intractable \cite{Huang2021PowerOfData,Huang2022MBSci,Vedran26ExpAdvObs,bokov2026luqpi}. 
Complementary experimental efforts have also begun to support the practical relevance of this paradigm by showing that learning can be performed directly on nontrivial quantum data acquired on quantum hardware~\cite{Huang2022SciQAdvExp,Cho2024QuantumExperimentalData}. 

Among the tools that make this setting practical, classical shadows provide a compact classical representation of quantum states obtained from randomized measurements, enabling the same measurement record to be reused for many downstream estimation tasks without full tomography \cite{Huang2020shadow,Elben2023RandomizedMeasurementToolbox,Zhang2021ExperimentalShadows}. More broadly, recent studies on learning from quantum measurement data include property prediction, quantum phase classification, adaptive measurement-based quantum state classification, learning directly from experimental quantum data, shadow-based modeling, and task-agnostic pretraining for quantum property estimation \cite{wang_predicting_2024,Cho2024QuantumExperimentalData,du_shadownet_2023,Tanji2026QPCHT,yao_shadowgpt_2024,Tang2024LLM4QPE,Bermejo2024QCNNSimulable,Li2025decision}.

Despite this progress, the assumptions of matched data distributions and abundant labels are rarely satisfied in realistic applications. 
In practice, labeled data are often available only in easier regimes such as numerically tractable parameter regions or well-controlled experiments.
By contrast, the target regime is typically the one in which reliable labels are hard to obtain, since certification is computationally expensive or experimental control and characterization are limited, leading to shifts in the underlying parameter distribution.
At the same time, the target data are subject to various sources of noise, including finite-shot noise, imperfect state preparation, readout errors and hardware noise \cite{Cao2025TrapsVQA}.
These mismatches naturally induce \emph{domain shift} between the data available for training and the data encountered in the target regime.
Although out-of-distribution generalization has been established for specific tasks in quantum-process learning and has also been demonstrated empirically in several many-body state and property prediction tasks \cite{Caro2023OODQuantumDynamics,Pereira2025OOD,Monaco2023MarginalPhaseOOD,Wu2024ShortRangeMTL,Rende2025FNQS}, extending learning protocols for quantum data to broader and more realistic forms of domain shift is an important step toward robust and practically useful quantum-data learning.

In this work, we address this challenge through \emph{unsupervised domain adaptation (UDA)} \cite{BenDavid2010LearningDifferentDomains,Ganin2016RV,pmlr-v37-long15,Wilson2020SurveyUDA}. 
Note that adversarial domain-adaptation ideas have previously been explored for quantum phase classification from ground-state data within classical framework \cite{PhysRevB.97.134109}. 
Here, we instead study the practically relevant setting in which labeled source-domain quantum data and unlabeled target-domain quantum data, represented by classical shadows, are available, with the two domains differing in the distribution of observed states and, potentially, in the effective relation between observed states and labels.
Since classical shadow provides classical data objects, adaptation and prediction can be carried out entirely within a classical computational pipeline once randomized measurements have been performed, without full state reconstruction or further quantum resources during training. This makes classical shadows a natural interface between quantum experiments and UDA.

We evaluate the proposed framework on two representative tasks under realistic domain shifts: quantum phase classification \cite{PhysRevB.97.134109,Huang2022MBSci,Bermejo2024QCNNSimulable} and entanglement classification \cite{Luo2023GME,Vintskevich2023entanglement,Koutn2023entanglement,Huang2025entanglement}.
In the phase-classification benchmarks, the shift arises from mismatched Hamiltonian-parameter regions together with imperfect ground-state preparation and hardware 
noise. In the entanglement benchmarks, the shift is induced by differences in system size, state-generation procedures and subsystem partitions. Through these experiments, we show that UDA on shadow data improves predictive performance relative to both source-only baselines and target-only unsupervised baselines under realistic distribution shifts. These results support UDA as a practical route toward more robust learning from imperfect quantum data.

The rest of this paper is organized as follows. In Sec.~\ref{sec:problem-setup}, we formulate the learning problem from the viewpoint of UDA for quantum data and explain how classical shadows naturally yield classical datasets suitable for this setting. In Sec.~\ref{sec:uda-method}, we present the proposed UDA pipeline for classical-shadow data. In Sec.~\ref{sec:numerics}, we report numerical results on the representative learning tasks considered in this work. Finally, Sec.~\ref{sec:conclusion} concludes with a discussion of the main findings and future directions. Additional numerical details are provided in the Appendix.

\section{Problem setup}
\label{sec:problem-setup}

\subsection{Unsupervised domain adaptation for unlabeled imperfect quantum data}

Let $\mathcal{H}_n := (\mathbb{C}^2)^{\otimes n}$ be the $n$-qubit Hilbert space and let $\cS(\mathcal{H}_n)$ denote the set of density operators on $\mathcal{H}_n$.
This work considers the task of predicting a classical label $Y\in\cY$ from an input quantum state.
Typical settings include classification $\cY=\{1,2,\dots,C\}$ and regression $\cY=\mathbb{R}$.
Conceptually, the label is assumed to be determined by an underlying ideal state $\rho^{\mathrm{ideal}}\in\cS(\mathcal{H}_n)$ through an unknown rule, such as a deterministic labeling map $\ell:\cS(\mathcal{H}_n)\to\cY$ satisfying $Y=\ell(\rho^{\mathrm{ideal}})$, or more generally a conditional distribution $Q(Y\mid \rho^{\mathrm{ideal}})$ of the joint input-label distribution $(\rho^{\mathrm{ideal}}, Y) \sim Q$.
For example, the task of classifying quantum phases of matter considers $\rho^{\mathrm{ideal}}$ as a ground state and $Y$ as its phase label (see Sec.~\ref{sec:numerics} for details).

In realistic situations, however, the observed states in the target regime typically include errors: even when a family of ideal states exists in theory, the actual states may be distorted by experimental imperfections, residual coherent errors, algorithmic imperfections and hardware errors in quantum devices.
As a result, the observed state and the label are generically governed by a different, unknown distribution,
\begin{align}
    (\rho^{\mathrm{t}}, Y^{\mathrm{t}})\sim Q^{\mathrm{t}} \neq Q,
\end{align}
where the superscripts/subscripts $\tgt$ denote the target data hereafter.
Hence, the task in a realistic scenario can be reduced to estimating the label $Y^{\tgt}$ from the observed state $\rho^{\tgt}$ following the unknown rule $Q^{\tgt}(Y^{\tgt}|\rho^{\tgt})$.
In a simple case, the observed state is related to the ideal state as
\begin{align}
  \rho^{\tgt} = \mathcal{E}\!\left(\rho^{\mathrm{ideal}}\right),
  \label{eq:target-channel}
\end{align}
where $\mathcal{E}$ is an unknown effective quantum channel that characterizes the imperfections.
In this case, the joint distribution of the observed state and the label is related to the ideal one as $Q^{\tgt}((\rho^{\tgt}, Y^t) \in A) = Q((\rho^{\ideal},Y) \in \mathcal{T}^{-1} (A))$ for any event $A \subset \cS(\mathcal{H}_n) \times \cY$, where $\mathcal{T}(\rho^{\ideal},y) := (\cE(\rho^{\ideal}),y)$.
In general, since imperfection is not necessarily represented by a fixed channel $\cE$ independent of the underlying state $\rho^{\ideal}$, we focus on the general target distribution $Q^{\tgt}(\rho^{\tgt}, Y^{\tgt})$.

Furthermore, the primary interest often lies in a regime in which many quantum states can be prepared and measured, while reliable labels are difficult or infeasible to obtain.
This difficulty can arise from the aforementioned imperfections or from the high computational cost for certifying labels due to classical intractability.
This is further exacerbated when the target system is challenging to control and characterize or available quantum-computing resources are insufficient to generate labels at scale.
Accordingly, learning from unlabeled imperfect target quantum data is of significant importance; this is the main motivation of our work.

More concretely, we assume access to the quantum state $\rho^{\tgt}$ without label $Y^{\tgt}$ with underlying distribution $(\rho^{\mathrm{t}}, Y^{\mathrm{t}}) \sim Q^{\mathrm{t}}$.
Although purely unsupervised learning approaches based only on the unlabeled target data are also conceivable \cite{Huang2022MBSci,morais2025distinguishing,khosrojerdi2025unsupervised}, they can be unreliable in the presence of imperfections.
Indeed, as demonstrated in the numerical experiments in Sec.~\ref{sec:numerics}, imperfect quantum data can substantially degrade the performance of such unsupervised baselines.

Even if labeled target data are unavailable, it is often realistic to have access to a related but differently distributed labeled dataset obtained in a more accessible regime.
Examples include regimes in which (i) the quantum experiment or quantum simulation is reliable and labels can be certified with moderate effort, or (ii) classical simulation remains tractable and labels can be computed.
That is, we have access to additional labeled data with the underlying distribution $(\rho^{\mathrm{s}}, Y^{\mathrm{s}}) \sim Q^{\mathrm{s}} \neq Q^{\mathrm{t}}$, which we refer to as the source data.
Throughout the manuscript, the superscripts/subscripts $\src$ will denote source data.
In addition to differences in the respective state's marginal distributions $Q_{\rho}^{\src}$ and $Q_{\rho}^{\tgt}$ of the source and the target domains across the accessible regimes, imperfections in the target data-generation process can alter the relation between an observed state and the underlying ideal state that determines the label.
Consequently, even when the same observed state $\rho$ appears in both domains, the conditional distribution can differ, $Q^{\tgt}(Y|\rho)\neq Q^{\src}(Y|\rho)$.
For example, if $\rho^{\tgt}=\mathcal{E}(\rho^{\mathrm{ideal}})$ while $\rho^{\src}=\rho^{\mathrm{ideal}}$, then the same observed $\rho$ can originate from different ideal states across domains.


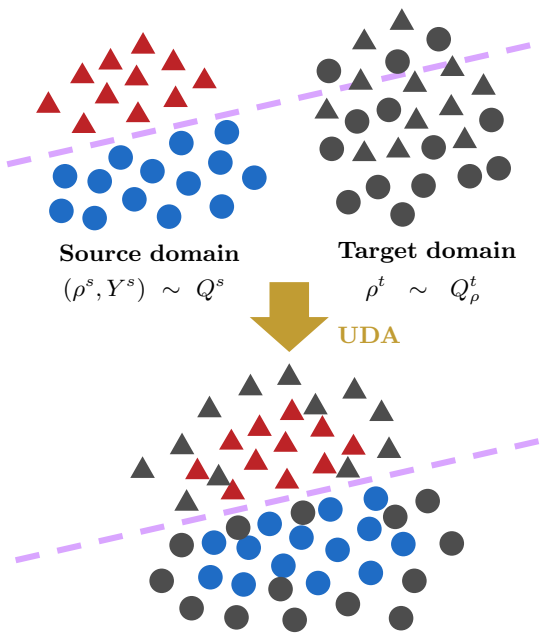
\begin{figure}[t]
\centering
\begin{tikzpicture}
\definecolor{lightgray}{HTML}{F4F4F4}
\definecolor{littlelightgray}{HTML}{ecececff}
\definecolor{pale}{HTML}{7ca3d4ff}
\definecolor{lightred}{HTML}{d8a2a2}
\definecolor{gold__}{HTML}{C19C33}
\node[text width=3cm, anchor=west, align=left ] at (-4.7,1.2){\textbf{Source domain}};
\node[text width=3cm, anchor=west, align=left ] at (-4.65,0.7){$(\rho^{s}, Y^{s})\sim Q^{s}$};
\node[text width=3cm, anchor=west, align=left ] at (-1.0,1.2){\textbf{Target domain}};
\node[text width=3cm, anchor=west, align=left ] at (-0.63,0.7){$\rho^{t}\sim Q^{t}_{\rho}$};
\node[anchor=center] (russell) at (-1.7,0.3)
{\centering\includegraphics[width=0.4\textwidth]{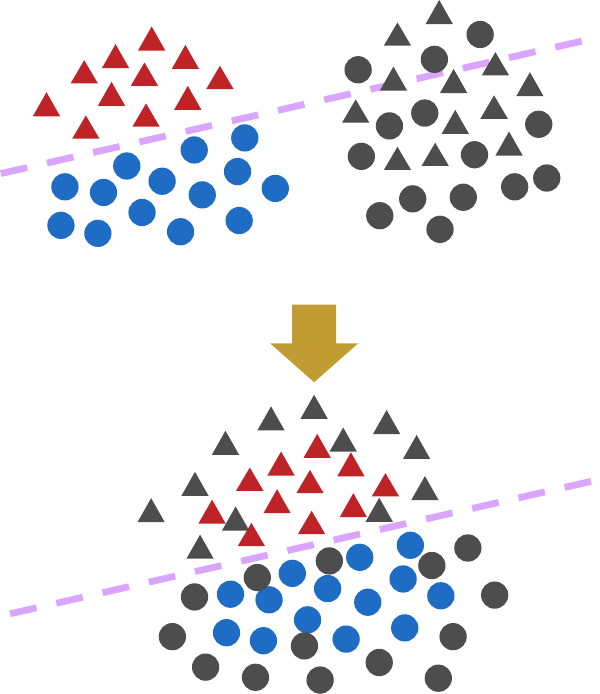}};
\node[text width=4cm, anchor=west, align=left ] at (-1.0,0.11){\textcolor{gold__}{\textbf{UDA}}};
\end{tikzpicture}
\caption{Schematic illustration of unsupervised domain adaptation (UDA) for quantum data. (Upper) The labeled source domain and the unlabeled target domain follow different distributions; as a consequence, the decision boundary learned on the source domain is generally misaligned with the target data. 
(Lower) A feature-extraction map is learned so that the resulting representation preserves label-relevant structure while reducing domain-specific discrepancies, enabling a decision rule that transfers from the source to the target domain. One such method is a domain-adversarial training as described in Sec.~\ref{sec:uda-method}.}
\label{fig:problem}
\end{figure}


The above setting, where a predictor is learned to perform well on the unlabeled target domain by exploiting labeled data from a related but different source domain, is known in machine learning as \textit{unsupervised domain adaptation} (UDA) \cite{BenDavid2010LearningDifferentDomains,Ganin2016RV,pmlr-v37-long15,Wilson2020SurveyUDA}.
In UDA, each data domain of source/target data is called source/target domain.
Fig.~\ref{fig:problem} schematically summarizes this setting.
In the original state space, the source and target domains can exhibit a substantial distribution mismatch, so that a predictor trained only on labeled source data may fail to generalize to the target domain.
UDA addresses this issue by learning a feature representation in which label-relevant structure is retained while domain-specific variation is suppressed, thereby improving transfer to unlabeled target data.
In particular, the present UDA setting is not restricted to the so-called covariate-shift $Q_{\rho}^{\src}\neq Q_{\rho}^{\tgt}$ under $Q^{\tgt}(Y|\rho)=Q^{\src}(Y|\rho)$, but it also allows for the concept shift $Q^{\tgt}(Y|\rho)\neq Q^{\src}(Y|\rho)$ \cite{MorenoTorres2012DatasetShift}.

\subsection{Classical dataset consisting of classical shadows}\label{subsec-csdata}

A practically relevant setting restricts the access of classical learning models to classical information obtained from quantum input states.
In particular, the setup does {\it not} assume full access to the quantum state~$\rho$ such as performing tomography.
Instead, each quantum state is measured locally to produce partial classical data, and then a classical learning algorithm is applied to the resulting classical datasets.

This framework is motivated by practical constraints. Maintaining a large collection of quantum states as a repeatedly accessible database would require quantum memory at scale, and end-to-end quantum learning models require coherent access to quantum data and nontrivial experimental overhead.
By contrast, storing classical information extracted from quantum states is compatible with current and near-term experimental workflows and offers an additional practical benefit: once a classical database is constructed, the same records can be reused for multiple downstream purposes, including estimating many different target properties even after data acquisition \cite{Huang2020shadow,huang2022learning,boyd2025high}.
While the learning stage remains classical, a potential quantum advantage or a practical quantum utility can still be expected when constructing the classical dataset from quantum data is prohibitive for purely classical methods. 
This expectation stems from the fact that some tasks are classically efficient with the aid of quantum data while classically hard without data~\cite{Huang2021PowerOfData,Vedran26ExpAdvObs,bokov2026luqpi,lewis2024improved}.
Such a regime may also arise even when the available quantum data are moderately imperfect.
Realizing this potential advantage in the actual learning task, however, still requires effective methods that can exploit such data.
The focus of this work is to provide such a method within the UDA framework.


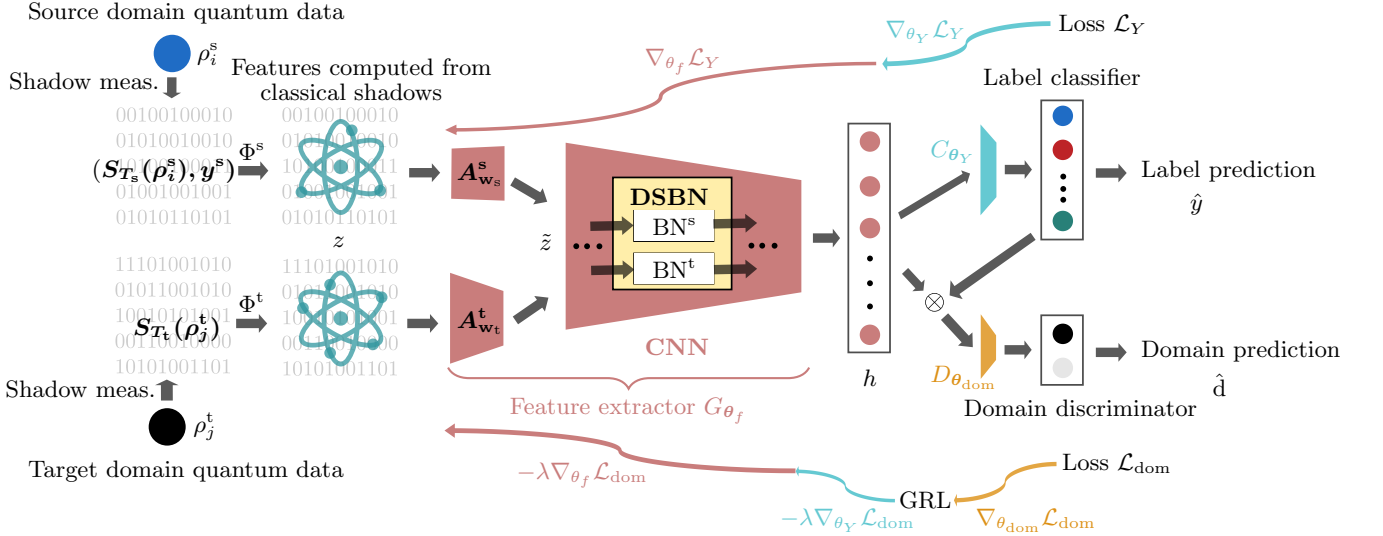
\begin{figure*}[t]
\centering
\begin{tikzpicture}
\definecolor{red__}{HTML}{C97D7C}
\definecolor{orange__}{HTML}{DD9117}
\definecolor{cian__}{HTML}{65C9D2}

\node[anchor=center] (russell) at (0,0)
{\centering\includegraphics[width=0.75\textwidth]{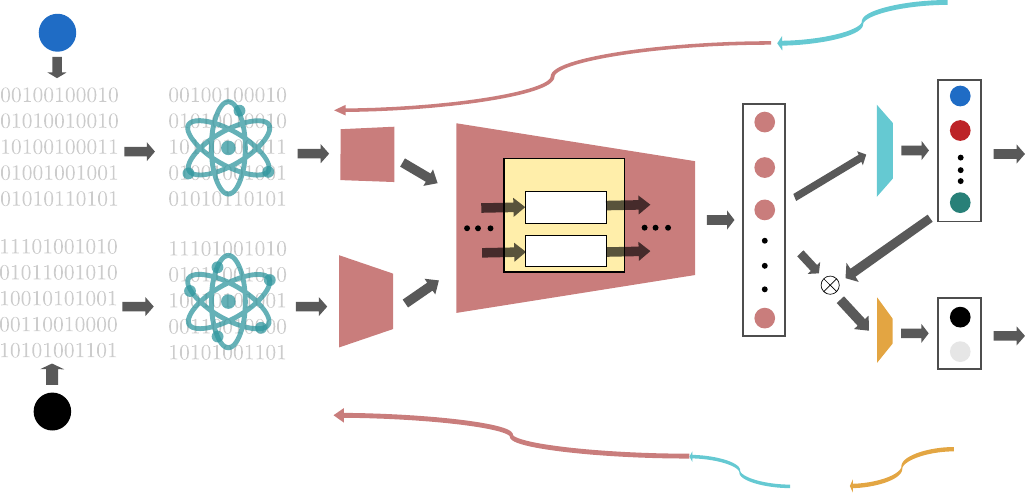}};

\modelgrid

\node[text width=5cm, anchor=west, align=left ] at (-5.75,2.8){$\rho^{\mathrm{s}}_i$};
\node[text width=5cm, anchor=west, align=left ] at (-5.8,-2.15){$\rho^{\mathrm{t}}_j$};
\node[text width=5cm, anchor=west, align=left ] at (-8.25,2.45){Shadow meas.};
\node[text width=5cm, anchor=west, align=left ] at (-8.3,-1.65){Shadow meas.};
\node[text width=5cm, anchor=west, align=left ] at (-8,3.35){Source domain quantum data};
\node[text width=5cm, anchor=west, align=left ] at (-8,-2.75){Target domain quantum data};
\node[text width=5cm, anchor=west, align=left ] at (-7.15,1.25){$(\bm{S_{T_{\mathrm{s}}}(\rho^{\mathrm{s}}_i), y^{\src})}$};
\node[text width=3cm, anchor=west, align=left ] at (-6.6,-0.85){$\bm{S_{T_{\mathrm{t}}}(\rho^{\mathrm{t}}_j)}$};
\node[text width=3cm, anchor=west, align=left ] at (-3.95,0.3){$z$};
\node[text width=3cm, anchor=west, align=left ] at (-1.2,0.3){$\tilde{z}$};
\node[text width=3cm, anchor=west, align=left ] at (-5.2,1.5){$\Phi^{\src}$};
\node[text width=3cm, anchor=west, align=left ] at (-5.2,-0.5){$\Phi^{\tgt}$};
\node[text width=5cm, anchor=west, align=left ] at (-5.3,2.6){Features computed from};
 \node[text width=3cm, anchor=west, align=left ] at (-4.91,2.3){classical shadows};
 \node[text width=5cm, anchor=west, align=left ] at (-1.5,-2.8){\textbf{\textcolor{red__}{$-\lambda\nabla_{\theta_f}\cL_{\dom}$}}};
 \node[text width=5cm, anchor=west, align=left ] at (0.2,2.7){\textbf{\textcolor{red__}{$\nabla_{\theta_f}\cL_{Y}$}}};
   \node[text width=5cm, anchor=west, align=left ] at (3.4,3.1){\textbf{\textcolor{cian__}{$\nabla_{\theta_Y}\cL_{Y}$}}};
 \node[text width=5cm, anchor=west, align=left ] at (2,-3.4){\textbf{\textcolor{cian__}{$-\lambda\nabla_{\theta_Y}\cL_{\dom}$}}};
   \node[text width=5cm, anchor=west, align=left ] at (4.6,-3.4){\textbf{\textcolor{orange__}{$\nabla_{\theta_{\dom}}\cL_{\dom}$}}};
\node[text width=5cm, anchor=west, align=left ] at (0.2,-1.1){\textbf{\textcolor{red__}{CNN}}};
\draw[
  red__,
  decorate,
  decoration={brace,mirror,amplitude=8pt,raise=2pt},
  thick
] (-2.3,-1.3) -- (2.5,-1.3)
node[midway,below=10pt] {Feature extractor $G_{\bm{\theta}_f}$};
\node[text width=5cm, anchor=west, align=left ] at (3.1,-1.5){$h$};
\node[text width=5cm, anchor=west, align=left ] at (0,0.9){\textbf{DSBN}};
\node[text width=5cm, anchor=west, align=left ] at (0.3,0.5){$\rm{BN}^{\src}$};
\node[text width=5cm, anchor=west, align=left ] at (0.3,-0.05){$\rm{BN}^{\tgt}$};

\node[text width=5cm, anchor=west, align=left ] at (-2.3,1.2){$\bm{A^{\src}}_{\mathbf{w}_{\src}}$};
\node[text width=5cm, anchor=west, align=left ] at (-2.3,-0.75){$\bm{A^{\tgt}}_{\mathbf{w}_{\tgt}}$};
\node[text width=3cm, anchor=west, align=left ] at (4.7,2.5){Label classifier};
\node[text width=5cm, anchor=west, align=left ] at (4.44,-1.9){Domain discriminator};
\node[text width=5cm, anchor=west, align=left ] at (4,-1.5){\textcolor{orange__}{$D_{\bm{\theta}_{\dom}}$}};
\node[text width=5cm, anchor=west, align=left ] at (4,1.5){\textcolor{cian__}{$C_{\bm{\theta}_{Y}}$}};
\node[text width=3cm, anchor=west, align=left ] at (6.8,1.25){Label prediction};
\node[text width=1cm, anchor=west, align=left ] at (7.45,0.8){$\hat{y}$};
\node[text width=3cm, anchor=west, align=left ] at (6.8,-1.15){Domain prediction};
\node[text width=1cm, anchor=west, align=left ] at (7.75,-1.6){$\hat{\dd}$};
\node[text width=3cm, anchor=west, align=left ] at (5.7,3.2){Loss $\cL_{Y}$};
\node[text width=3cm, anchor=west, align=left ] at (5.75,-2.65){Loss $\cL_{\dom}$};
\node[text width=3cm, anchor=west, align=left ] at (3.58,-3.13){GRL};
\end{tikzpicture}
\caption{Learning pipeline for unsupervised domain adaptation from classical shadow datasets based on CDAN.
The input is the labeled source data $(S_{T_{\mathrm{s}}}(\rho^{\mathrm{s}}_i), y)\in\mathcal{D}^{\src}_{N_{\src}}$ or the unlabeled target data $S_{T_{\mathrm{s}}}(\rho^{\mathrm{s}}_j) \in \mathcal{D}^{\tgt}_{N_{\tgt}}$, each consisting of Pauli classical shadow.
A feature map $\Phi^{\dd}$ converts each shadow record to an input feature tensor $z$ whose size may depend on the domain $d$.
The feature extractor $G_{\bm{\theta}_f}$ consists of an input-size adapter $A^{\dd}_{\mathbf{w}_{\dd}}$ to convert $z$ to a fixed-shape input $\tilde{z}$ followed by a CNN that takes $\tilde{z}$ as input.
The CNN includes DSBN layers $\mathrm{BN}^{\dd}$.
The latent feature $h = G_{\bm{\theta}_f}(z)$ is converted to the label probability distribution via a label classifier $C_{\bm{\theta}_Y}$ to give a label prediction $\hat{y}$ by taking the argmax.
The Kronecker product of $h$ and the label probability distribution is fed into a domain discriminator $D_{\bm{\theta}_\dom}$, which outputs a prediction of which domain the given data belongs to.
While the feature extractor and the label classifier are trained to minimize the label classification loss $\cL_Y$ using labeled source examples, they are simultaneously trained to maximize the domain discrimination loss $\cL_\dom$ via GRL.
The domain discriminator is adversarially trained to minimize $\cL_\dom$.
This way, the label-conditional latent feature distributions over the two domains are made similar while keeping the class-label distinguishability.}
\label{fig:model}
\end{figure*}



Among several measurement protocols, a particularly simple and experimentally feasible option is the Pauli classical shadow \cite{Huang2020shadow}, where single-qubit random Pauli measurements are performed to efficiently store classical snapshots of quantum data.
Although Clifford classical shadows~\cite{Huang2020shadow} or entangled measurements across multiple copies of the state~\cite{Huang2021QAdvPRL.126.190505, Huang2022SciQAdvExp} offer stronger theoretical guarantees, this work adopts Pauli classical shadows as the data-extraction method due to their ease of implementation and compatibility with near-term experimental platforms.

The protocol proceeds as follows.
Given an $n$-qubit state $\rho$, $T$ independent measurement shots are collected.
In Pauli classical shadow \cite{Huang2020shadow}, a random measurement setting is chosen for each shot $j\in\{1,\dots,T\}$:
\begin{align}
 \mathbf{P}_j = (P_{1,j},\dots,P_{n,j})\in\{X,Y,Z\}^n.
\end{align}
Each qubit labeled $i$ is then measured in the eigenbasis of $P_{i,j}$, and the corresponding outcome bit $b_{i,j}\in\{0,1\}$ is recorded.
The eigenstate of each qubit after measurement in the chosen setting can be represented as
\begin{align}
\ket{s_{i,j}} \in \{\ket{0}, \ket{1}, \ket{+}, \ket{-}, \ket{\mathrm{i}+}, \ket{\mathrm{i}-}\},
\end{align}
where $\ket{0},\ket{1}$ are $Z$-eigenstates, $\ket{\pm}$ are $X$-eigenstates, and $\ket{\mathrm{i}\pm}$ are $Y$-eigenstates.
Hence, the classical snapshot of $j$-th shot with the measurement setting $\mathbf{P}_j$ can be encoded as the product state as $\ket{s_j} := \bigotimes_{i=1}^n \ket{s_{i,j}}$.
We denote the resulting classical shadow data by
\begin{align}
  S_T(\rho)
  &:=
  \bigl\{(\mathbf{P}_j,\mathbf{b}_j)\bigr\}_{j=1}^T \nonumber \\
  &\equiv
  \bigl\{\ket{s_{i,j}} : i\in\{1,\dots,n\},\; j\in\{1,\dots,T\}\bigr\},
  \label{eq:shadow-data}
\end{align}
where $\mathbf{b}_j:=(b_{1,j},\dots,b_{n,j})$.
The key point is that $S_T(\rho)$ is a purely classical object that can be stored and reused.
Moreover, by post-processing the classical snapshots, many different properties of $\rho$ can be estimated from the same shadow dataset, without reconstructing $\rho$ itself.

Within this setup, our goal is hence summarized as classical learning of the prediction rule $Q^{\tgt}(Y^{\tgt}|\rho^{\tgt})$ from the unlabeled target-domain classical data 
\begin{align}
  \mathcal{D}^{\mathrm{t}}_{N_{\tgt}}
  :=
  \bigl\{S_{T_{\mathrm{t}}}(\rho^{\mathrm{t}}_j)\bigr\}_{j=1}^{N_{\mathrm{t}}},
  \qquad \rho^{\mathrm{t}}_j \stackrel{\mathrm{i.i.d.}}{\sim} Q_{\rho}^{\mathrm{t}},
  \label{eq:target-dataset}
\end{align}
with the aid of the labeled source-domain classical data
\begin{align}
  \mathcal{D}^{\mathrm{s}}_{N_{\src}}
  :=
  \bigl\{(S_{T_{\mathrm{s}}}(\rho^{\mathrm{s}}_i),\, y^{\mathrm{s}}_i)\bigr\}_{i=1}^{N_{\mathrm{s}}},
  \qquad (\rho^{\mathrm{s}}_i,y^{\mathrm{s}}_i)\stackrel{\mathrm{i.i.d.}}{\sim} Q^{\mathrm{s}},
  \label{eq:source-dataset}
\end{align}
where, in general, the joint distributions differ,
\[
(\rho^{\mathrm{s}},Y^{\mathrm{s}})\sim Q^{\mathrm{s}},
\qquad
(\rho^{\mathrm{t}},Y^{\mathrm{t}})\sim Q^{\mathrm{t}},
\qquad
Q^{\mathrm{s}}\neq Q^{\mathrm{t}}.
\]
Accordingly, this work adopts the UDA viewpoint to address learning from unlabeled Pauli classical shadows of imperfect quantum data.
The next section specifies how classical-shadow-derived features are constructed and how UDA objectives are instantiated in the learning pipeline.

\section{Unsupervised domain adaptation methods for classical shadow dataset}
\label{sec:uda-method}

This section presents the learning methods for UDA using Pauli classical shadow datasets.
For concreteness, we formulate the pipeline focusing on the classification tasks, although it can be generalized to the regression.
We first introduce the overall pipeline of the method and then describe its main components in detail.
Since target labels are unavailable in UDA, we also introduce a label-free model-selection strategy for hyperparameter tuning later in Sec.~\ref{subsec:model-selection}.

\subsection{Whole pipeline}
\label{subsec:whole-pipeline}

Fig.~\ref{fig:model} summarizes the whole pipeline with overall data flow.
The classical shadow of each quantum data is used to extract an input feature tensor.
This feature tensor is processed by a neural feature extractor trained jointly with a label classifier and a domain discriminator within the conditional domain adversarial network~(CDAN) framework \cite{Long2018CDAN}.
The adversarial objective promotes representations that remain discriminative for the prediction task while remaining approximately invariant to domain shifts between the source and target datasets.

More concretely, given a classical shadow record $S_{T_{\dd}}(\rho^{\dd})$ from each domain $d \in \{\src, \tgt\}$, we first compute a feature tensor $z=\Phi^{\dd}(S_{T_{\dd}}(\rho^{\dd}))$ by applying a classical post-processing map $\Phi^{\dd}$ as detailed in Sec.~\ref{subsec:shadow-features}.
Here, the map $\Phi^{\dd}$ may be domain dependent in general if each domain has distinct size of the underlying quantum system.
A domain-dependent input-size adapter $A_{\mathbf{w}_{\dd}}^{\dd}$ is then applied to obtain $\tilde z = A_{\mathbf{w}_{\dd}}^{\dd} z$ to resolve the size difference if needed (see Sec.~\ref{subsec:dsbn-adapter}).
The resulting representation is subsequently processed by a convolutional neural network (CNN) $\widetilde{G}^{\dd}_{\theta_{\mathrm{CNN}}^{\dd}}$ to produce the latent embedding, whose dependence on $d$ comes from the domain-specific batch normalization (DSBN) \cite{Chang2019DSBN} layers of CNN.
As described in Sec.~\ref{subsec:dsbn-adapter}, each batch normalization layer $\mathrm{BN}$ is replaced by a domain specific one $\mathrm{BN}^{\dd}$ to accommodate for the statistical difference between the domains.
The resulting feature extractor $G^{\dd}_{\bm{\theta}_f}$ reads
\[
h = G_{\bm{\theta}_f}^{\dd}(z) = \widetilde{G}^{\dd}_{\bm{\theta}_{\mathrm{CNN}}^{\dd}}(A_{\mathbf{w}_{\dd}}^{\dd}z),
\]
where $\bm{\bm{\theta}}_f$ denotes the whole trainable parameters of the feature extractor including the input-size adapters and CNNs for both domains.

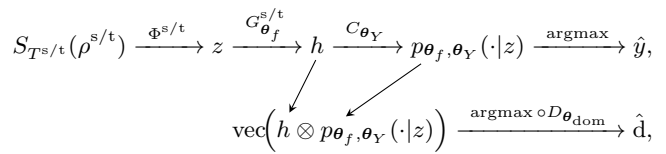
\begin{figure}[h]
\begin{flushleft}
\begin{tikzpicture}[>=stealth]

\node[anchor=east] (eq1) at (0,0) {$\displaystyle
S_{T^{\src/\tgt}}(\rho^{\src/\tgt})
\xrightarrow{\ \Phi^{\src/\tgt}\ } z
\xrightarrow{\ G_{\bm{\theta}_f}^{\src/\tgt}\ } h
\xrightarrow{\ C_{\bm{\theta}_Y}\ }
p_{\bm{\theta}_f,\bm{\theta}_Y}(\cdot| z)
\xrightarrow{\ \argmax\ } \hat y,
$};

\node[anchor=east] (eq2) at (0,-1.25) {$\displaystyle
\mathrm{vec}\!\Bigl(h\otimes p_{\bm{\theta}_f,\bm{\theta}_Y}(\cdot| z)\Bigr)
\xrightarrow{\ \argmax \circ D_{\bm{\theta}_{\dom}}\ } \hat \dd,
$};

 \draw[->] (-4.6,-0.35) -- (-4.95,-1.05);
\draw[->] (-3.2,-0.4) -- (-4.2,-1.1);

\end{tikzpicture}
\end{flushleft}

\caption{The forward mapping of the whole pipeline.}
\label{fig:process}
\end{figure}

The label classifier $C_{\bm{\bm{\theta}}_Y}$ consists of a single linear layer with trainable paramters $\bm{\bm{\theta}}_Y$ and the softmax layer to output class probability distribution
\[
(p_{\bm{\bm{\theta}}_f,\bm{\bm{\theta}}_Y}(y| z))_y= C_{\bm{\bm{\theta}}_Y}(h).
\]
The label loss $\cL_{Y}$ is evaluated only on the labeled source examples.
In CDAN, the input $u=\mathrm{vec} \bigl(h\otimes p_{\bm{\bm{\theta}}_f,\bm{\bm{\theta}}_Y}(\cdot| z)\bigr)$ is fed into the domain discriminator $D_{\bm{\bm{\theta}}_{\dom}}$, where $\otimes$ denotes the Kronecker product and $\bm{\theta}_{\dom}$ denotes the trainable parameters.
The discriminator also consists of a single linear layer and the softmax layer to output the probability
\begin{align*}
    q_{\bm{\theta}_f, \bm{\theta}_{Y}, \bm{\theta}_{\dom}}(d|z) =
    \begin{cases}
    D_{\bm{\theta}_{\dom}}(u) \quad &(d=\src)\\
    1 - D_{\bm{\theta}_{\dom}}(u) \quad &(d=\tgt).
    \end{cases}
\end{align*}
The discriminator is trained on both source and target examples so that the head paramters $\bm{\bm{\theta}}_{\dom}$ are updated to decrease the domain loss $\cL_{\dom}$, while
the feature extractor $G^{\dd}_{\bm{\bm{\theta}}_f}$ is updated to increase the domain loss $\cL_{\dom}$ via the gradient reversal layer (GRL) as described in Sec.~\ref{subsec:cdan}.
The resulting forward mapping is depicted in Fig.~\ref{fig:process}.


\subsection{Input feature tensor from classical shadows}
\label{subsec:shadow-features}

Each classical shadow record $S_T(\rho)$ is first mapped into an input representation $z$ for the downstream neural network.
Classical shadows provide sample-efficient estimates of expectation values for a large set of local observables through classical post-processing.
Accordingly, the representation $z$ can be regarded as an input feature tensor constructed from estimated local observables, where the specific choice of observables is guided by the learning task and the structural properties of the underlying quantum data.

More specifically, let $\{O_\alpha\}_\alpha$ be a collection of observables.
From $S_T(\rho)$, we can compute estimators $\widehat{\langle O_\alpha\rangle}_\rho$ for $\langle O_\alpha\rangle_\rho := \Tr[O_\alpha \rho]$.
Then, we define a feature map
\[
z = \Phi^{\dd}(S_{T_{\dd}}(\rho^{\dd})) := \bigl(\widehat{\langle O^{\dd}_\alpha\rangle}_\rho\bigr)_\alpha 
\]
with each observable set $\{O^{\dd}_\alpha\}_\alpha$ for domain $\dd$.
Those observables should be chosen so that $z$ captures label-relevant information while remaining feasible to estimate with the available number of shots $T_{\dd}$.

In particular, for quantum data whose system is defined on a lattice, it is natural to organize the features as a tensor aligned with the lattice \cite{Huang2022MBSci}. 
Let $\Lambda = \{1,...,L_1\} \otimes \cdots \otimes \{1,...,L_{D}\}$ denote the $D$-dimensional lattice of size $L_1\times \dots \times L_D$, and write each site as $\bm{i} \in \Lambda$.
For simplicity, we focus on features constructed from two-site reduced density matrices. The extension to general $r$-site reduced density matrices is straightforward.

For each site $\bm{i} \in \Lambda$, choose an ordered subset
$\Lambda_{\bm{i}} = \{\bm{j}_{\bm{i}}^{(1)},\dots,\bm{j}_{\bm{i}}^{(M)}\} \subset \Lambda \backslash \{\bm{i}\}$.
We assume that $|\Lambda_{\bm{i}}| = K$ or $0$ with a fixed number $K$.
For example, for a 1D lattice $\Lambda = \{1,\dots, L\}$ with the open boundary condition, one may take up to the $k$-nearest-neighbor sites 
\begin{align}
 \Lambda_i^{(k)}
 =\begin{cases}
     \{i + 1, \dots, i+k\} \quad & (i=1,\dots,L - k)\\
     \emptyset \quad & (i= L - k + 1, \dots, L),
 \end{cases}\label{eq:lmd-i-k}
\end{align}
where $K=k$ holds in this case.

For each single-qubit Pauli operator $P \in \{I,X,Y,Z\}$, let $P_{\bm{i}}$ denote the $|\Lambda|$-qubit operator with weight one that acts as $P$ on site $\bm{i}$ and as the identity on every other site.
We also fix an ordering
\begin{align}
&\{(P,Q)\in \{I,X,Y,Z\}^{2} : (P,Q)\neq (I,I)\}\notag \\
=&
\{(P^{(a)},Q^{(a)})\}_{a=1}^{15}.
\end{align}
Then, for each pair $(\bm{i},\bm{j})$ with $\bm{j}\in \Lambda_{\bm{i}}$, we estimate the $15$ nontrivial two-site Pauli expectation values
\begin{align}
\hat{c}_{\bm{i},\bm{j}}^{(a)}
:=
\widehat{\langle P^{(a)}_{\bm{i}} Q^{(a)}_{\bm{j}} \rangle}_{\rho},
\qquad
a=1,\dots,15.
\end{align}
These coefficients completely characterizes the two-site reduced density matrix of sites $\bm{i},\bm{j}$ through its Pauli-basis expansion.
For each $\bm{i}$ and $m=1,\dots,K$, define the $15$-dimensional vector
\begin{align}
\hat{\bm{c}}_{\bm{i}}^{(m)}
:=
\bigl(
\hat{c}_{\bm{i},\,\bm{j}_{\bm{i}}^{(m)}}^{(1)},
\dots,
\hat{c}_{\bm{i},\,\bm{j}_{\bm{i}}^{(m)}}^{(15)}
\bigr)^{\top}
\in \mathbb{R}^{15}.
\end{align}
We then define the feature vector at site $\bm{i}$ by concatenating these vectors over all $\bm{j}\in \Lambda_{\bm{i}}$:
\begin{align}
\bm{z}_{\bm{i}}
:=
\bigl(
(\hat{\bm{c}}_{\bm{i}}^{(1)})^{\top},
\dots,
(\hat{\bm{c}}_{\bm{i}}^{(M)})^{\top}
\bigr)^{\top}
\in \mathbb{R}^{15M}.
\end{align}
Equivalently, the feature tensor entries can be written explicitly as
\begin{align}
&z_{\bm{i},\,l}
:=
\hat{c}_{\bm{i},\,\bm{j}_{\bm{i}}^{(m)}}^{(a)}
\nonumber\\
&(l = 15(m-1)+a,
m=1,\dots,K,\;
a=1,\dots,15).
\end{align}
Thus, the full feature tensor is
\begin{align}
z = (z_{\bm{i},l})
\in
\mathbb{R}^{L_{1}\times \cdots \times L_{D}\times 15K}.
\end{align}
This tensor can be fed directly into a $D$-dimensional CNN, with $\bm{i}$ treated as the spatial index and $l$ as the input-channel index.
We can also fed the feature tensor $z$ to one-dimensional CNN by vectorizing $\bm{i}$.
This representation allows the CNN to learn nonlinear functions of local reduced states involving correlations captured by~$\Lambda_{\bm{i}}$.
Especially, for $D=1$ with $\Lambda_i^{(k)}$ defined by Eq.~\eqref{eq:lmd-i-k}, we denote the resulting feature map $S_T(\rho)\mapsto z$ by $\Phi^1_k$, which can take into account correlations up to $k$-nearest-neighbor.
In our numerics for one dimensional systems in Sec.~\ref{subsec:qpt}, we use this family of local reduced-density feature maps with an appropriate choice of $k$ for each task.

\subsection{Domain-specific batch normalization and input-size adapter}
\label{subsec:dsbn-adapter}

Our CNN feature extractor contains batch normalization (BN) layers, which normalize intermediate activations using batch statistics and learned affine parameters.
In UDA, sharing BN layers across domains can be suboptimal because activation statistics can differ substantially between the source and target domains.
To account for this, we employ DSBN \cite{Chang2019DSBN}, which maintains separate BN parameters and running statistics for the source and target domains.
Concretely, each BN layer is replaced by a pair $\mathrm{BN}^{\src}$ and $\mathrm{BN}^{\tgt}$ with independent affine parameters and running moments, and the appropriate branch is selected according to the domain of each input.
The remaining convolutional weights are shared.

In addition, some datasets involve different input dimensions across domains, for example, due to different feature constructions or different numbers of local observables.
To handle this setting within a unified architecture, we introduce a trainable linear input-size adapter $A^{\src/\tgt}_{\mathbf{w}_{\src/\tgt}}$ that maps each domain input into a common dimension:
\[
\tilde z =
\begin{cases}
A^{\src}_{\mathbf{w}_{\src}} z, & (\text{$z$ from the source domain}),\\
A^{\tgt}_{\mathbf{w}_{\tgt}} z, & (\text{$z$ from the target domain}),
\end{cases}
\]
where $\tilde z$ has a fixed dimension shared by both domains.
When the source and target inputs already have the same dimension, we set $A^{\src}_{\mathbf{w}_{\src}}=A^{\tgt}_{\mathbf{w}_{\tgt}}=I$ and omit this adapter.
Fig.~\ref{fig:model} includes this input-adaptation stage as an explicit component when needed.

With DSBN and input-size adapter, the feature extractor becomes domain conditioned.
One may decompose $\bm{\bm{\theta}}_f$ into shared CNN parameters $\bm{\theta}_{\mathrm{CNN}}^{\mathrm{sh}}$ and the domain-specific parts as
$\bm{\theta}_f = (\bm{\theta}_{\mathrm{CNN}}^{\mathrm{sh}}, \bm{\theta}_{\mathrm{CNN}}^{\mathrm{BN}^{\src}}, \bm{\theta}_{\mathrm{CNN}}^{\mathrm{BN}^{\tgt}}, \mathbf{w}_{\src}, \mathbf{w}_{\tgt})$.
For brevity, we use the shorthand $G_{\bm{\bm{\theta}}_f}^{\src/\tgt}$ for the feature extractor and do not expand domain-dependent parameters explicitly in the subsequent notation.

\subsection{Conditional domain adversarial network}
\label{subsec:cdan}



We next describe the details of CDAN.
In the original domain-adversarial neural network (DANN) formulation for UDA \cite{Ganin2016RV},
the discriminator directly takes the latent feature $h$ as input, thereby encouraging alignment of the latent marginal distributions of both domains.
However, such marginal alignment can be insufficient when the data distribution has complex multi-modal structures~\cite{Long2018CDAN}.
CDAN was invented to address this issue by conditioning the discriminator on the predicted class probabilities~\cite{Long2018CDAN}.
Concretely, CDAN feeds
\[
u = \mathrm{vec}\!\Bigl(h \otimes p_{\bm{\bm{\theta}}_f,\bm{\bm{\theta}}_Y}(\cdot|z)\Bigr),
\]
into the discriminator $D_{\bm{\theta}_{\dom}}$ to take into account the label-conditional latent distribution.
Furthermore, we employ the entropy conditioning to suppress the contribution of uncertain predictions.
The uncertainty of a probability distribution $(p(y))_y$ is quantified by the entropy $H(p) = - \sum_y p(y)\log p(y)$, and the entropy-aware weight $w(z) = 1 + e^{-H(p_{\bm{\bm{\theta}}_f,\bm{\bm{\theta}}_Y}(\cdot|z))}$ is applied.
Hence, the domain loss is defined as
\begin{align}
&\cL_{\dom}(\bm{\bm{\theta}}_f,\bm{\bm{\theta}}_Y,\bm{\bm{\theta}}_{\dom})\notag\\
=&
\mathbb{E}_{\rho^{\src} \sim Q_{\rho}^{\src}}
\Bigl[
-w(z^{\src})\log D_{\bm{\bm{\theta}}_{\dom}}\bigl(u^{\src}\bigr)
\Bigr]\notag\\
&+
\mathbb{E}_{\rho^{\tgt}\sim Q_{\rho}^{\tgt}}
\Bigl[
-w(z^{\tgt}) \log \left(1 - D_{\bm{\bm{\theta}}_{\dom}}\bigl(u^{\tgt}\bigr)\right)
\Bigr],
\end{align}
while the label loss $\cL_Y$ is computed only on labeled source examples:
\[
\cL_Y(\bm{\bm{\theta}}_f,\bm{\bm{\theta}}_Y)
=
\mathbb{E}_{(\rho^{\src},y^{\src})\sim Q^{\src}}
\Bigl[
-\log p_{\bm{\bm{\theta}}_f,\bm{\bm{\theta}}_Y}\bigl(y=y^{\src}\,|\,z\bigr)
\Bigr].
\]
The overall training objective follows the standard adversarial formulation, where
the discriminator minimizes $\cL_{\dom}$, while the feature extractor and the label classifier minimize $\cL_Y$ and simultaneously confuse the discriminator.
With a trade-off hyperparameter $\lambda>0$ and $\cL(\bm{\bm{\theta}}_f,\bm{\theta}_Y,{\bm{\theta}}_{\dom}):=\cL_Y(\bm{\bm{\theta}}_f,\bm{\bm{\theta}}_Y) - \lambda \cL_{\dom}(\bm{\bm{\theta}}_f,\bm{\bm{\theta}}_Y,\bm{\bm{\theta}}_{\dom})$, this corresponds to the minimax problem
\begin{align}
\min_{(\bm{\bm{\theta}}_f,\bm{\bm{\theta}}_Y)} &\cL(\bm{\bm{\theta}}_f,\bm{\theta}_Y,{\bm{\theta}}_{\dom}),\\
 \max_{{\bm{\theta}}_{\dom}} \ &\cL(\bm{\bm{\theta}}_f,\bm{\theta}_Y,{\bm{\theta}}_{\dom}).
\end{align}
In practice, this is implemented by the backpropagation with a GRL inserted between the discriminator and the rest of the network.
Writing the updates in a stochastic gradient descent (SGD) form with learning rate $\eta>0$ using mini-batch estimates $\widehat{\cL}_Y,\widehat{\cL}_{\dom}$, we have
\begin{align}
    \bm{\theta}_{\dom} & \leftarrow \bm{\theta}_{\dom} - \eta \nabla_{\bm{\theta}_{\dom}}\widehat{\cL}_{\dom},\\
    \bm{\theta}_f & \leftarrow \bm{\theta}_f - \eta\Bigl(\nabla_{\bm{\theta}_f}\widehat{\cL}_Y - \lambda \nabla_{\bm{\theta}_f}\widehat{\cL}_{\dom}\Bigr),\\
    \bm{\theta}_Y &\leftarrow \bm{\theta}_Y - \eta (\nabla_{\bm{\theta}_Y}\widehat{\cL}_Y - \lambda \nabla_{\bm{\theta}_Y}\widehat{\cL}_{\dom}\big).
\end{align}


\subsection{Model selection without target labels}
\label{subsec:model-selection}

Hyperparameter tuning and epoch selection are nontrivial in the UDA setting, since target labels are unavailable and standard cross-validation on the target domain cannot be applied.
Consequently, model selection must rely on label-free signals computable from unlabeled target data, possibly together with labeled source data.
A variety of such criteria have been proposed in the literature, including source domain-based methods \cite{Ganin2015SourceRisk,Sugiyama2007IWCV,You2019DEV,Ganin2016RV} and target domain-based methods \cite{Morerio2017Entropy,Musgrave2022InfoMax,Saito2021SND,NEURIPS2024_f50cebc2}.
Among these, we employ two of the target domain-based methods below.


We proceed as follows.
For each candidate hyperparameter configuration, we train a CDAN model using labeled source data and unlabeled target data.
Let $\cM_{1},\dots,\cM_{K}$ denote the resulting trained models.
For each model $\cM_{m}$ and each unlabeled target example $z^{\tgt}$, we compute the output label prediction probability distribution $p_{\cM_m}(y|z^{\tgt})$.
We then evaluate the following label-free criteria on the target domain data:

\begin{itemize}
    \item \textbf{InfoMax}~\cite{Musgrave2022InfoMax, NEURIPS2024_f50cebc2}.
    Information Maximization (InfoMax) uses input-output mutual information maximization as a matric \cite{NIPS1991_a8abb4bb_mutual_info,Musgrave2022InfoMax, NEURIPS2024_f50cebc2}.
    InfoMax balances low conditional entropy on individual target examples with high marginal entropy of the average prediction, which discourages degenerate solutions that collapse all predictions to a single class.
    Concretely, define the target predictive entropy
    \[
    H_{\mathrm{tgt}}(\cM_{m}) = \mathbb{E}_{\rho^{\tgt}\sim Q_{\rho}^{\tgt}} \bigl[ H\bigl(p_{\cM_{m}}(\cdot|z^{\tgt})\bigr) \bigr],
    \]
    and the marginal entropy of the average prediction
    \[
    H\Bigl(\mathbb{E}_{\rho^{\tgt}\sim Q_{\rho}^{\tgt}}[p_{\cM_m}(\cdot|z^{\tgt})]\Bigr).
    \]
    The InfoMax score is then an empirical estimation of
    \[
    \quad \qquad I_{\mathrm{tgt}}(\cM_{m})
    =
    H\Bigl(\mathbb{E}_{\rho^{\tgt}\sim Q_{\rho}^{\tgt}}[p_{\cM_m}(\cdot|z^{\tgt})]\Bigr)
    -
    H_{\mathrm{tgt}}(\cM_{m}),
    \]
    and models with larger values of $I_{\mathrm{tgt}}(\cM_{m})$ are preferred.

    \item \textbf{EnsV}\cite{NEURIPS2024_f50cebc2}.
    Ensemble-based Validation (EnsV) selects a model that best agrees with the consensus behavior across candidate models on the unlabeled target data \cite{NEURIPS2024_f50cebc2}.
    In EnsV, we treat the ensemble-average prediction
    \[
    \bar{p}(y|z^{\tgt}) = \frac{1}{M} \sum_{m=1}^{M} p_{\cM_m}(y|z^{\tgt})
    \]
	  as a virtual teacher and select the model with the largest agreement with $\bar{p}$.
	  Intuitively, this favors models that match the stable predictions shared across configurations, while reducing sensitivity to idiosyncratic behaviors of any single run.
	  We can use an arbitrary quantification of the agreement between each candidate model and $\bar{p}$ depending on the purpose.
    A simple implementation uses the average inner product
    \[
    A(\cM_{m}) = \mathbb{E}_{\rho\sim Q_{\rho}^{\tgt}} \bigl[ \langle p_{\cM_m}(\cdot|z^{\tgt}), \bar{p}(\cdot|z^{\tgt}) \rangle \bigr].
    \]
    For classifications, we can use a classification score metric such as macro-F1 score regarding $\bar{p}$ as the ground truth, as we do in the numerics in Sec.~\ref{sec:numerics}.
    
\end{itemize}

\section{Numerical results}
\label{sec:numerics}

This section evaluates the UDA pipeline introduced in Sec.~\ref{sec:uda-method} on two quantum-data classification tasks: classification of quantum phases of matter and classification of entangled states. 

We briefly describe the common settings used in the following numerical experiments.
We first generate the underlying quantum states by randomly sampling $N_{\src}=400$ labeled source samples and $N_{\tgt}=800$ unlabeled target samples, with the numbers of samples balanced across classes, and then keep these quantum states fixed throughout $10$ trials. In each trial, the target samples are randomly split evenly into two subsets: $400$ samples are used as unlabeled target-train data for adaptation, and the remaining $400$ samples are held out as unseen target data for score evaluation. Thus, the subset of target samples used for adaptation varies across trials. In each trial, the entire pipeline is executed end-to-end, including (i) generating classical-shadow measurement data, (ii) constructing shadow-based classical features, (iii) training models, and (iv) selecting a model without target labels using the methods in Sec.~\ref{subsec:model-selection}.

The reported performance metrics are the median, the mean and standard deviation of the macro-F1 score of the unseen target data over these $10$ trials.
In our numerics, we use a one-dimensional CNN with DSBN for the feature extractor, and linear heads for both class prediction and domain discrimination in the UDA pipeline specified in Sec.~\ref{sec:uda-method}.
The detailed implementation is given in Appendix~\ref{app:network-architectures}.
We use the Adam optimizer \cite{Kingma:2015Adam} to train the models.
The hyperparameter grids are
\begin{align}
  E &\in \{200,300,400,500,600,700,800,900,1,000\}, \nonumber\\
  B &\in \{20,50,80\}, \nonumber\\
  \eta &\in \{10^{-6}, 5\times 10^{-6}, 10^{-5}, 5\times 10^{-5}, 10^{-4}\}, \label{eq:hpo_grid_common}
\end{align}
where $E, B$ and $\eta$ are the number of epochs, the mini-batch size, and the learning rate, respectively.
For CDAN, the domain-loss weight $\lambda$ is additionally swept over
\begin{equation}
  \lambda \in \{0.5, 1.0, 1.5\}.
\end{equation}


The target-domain quantum data include state-preparation imperfections, while the source-domain quantum data are taken to be clean.
State-preparation imperfections are introduced in a task-dependent manner and are specified for each task.
We compare the UDA method with the following two baselines that do not perform domain adaptation:

\begin{enumerate}[label=(\roman*), align=left]
    \item \textit{Source-only empirical risk minimization (ERM).}
    This baseline corresponds to training the same CNN feature extractor and a linear classifier only on labeled source data, and then applying the resulting model to the target data without adaptation.
For model selection, we use cross-validation on the source domain.
More precisely, for each hyperparameter configuration and candidate number of epochs in \eqref{eq:hpo_grid_common}, we select the best setting using three-fold cross-validation on the source dataset, and then refit the model on the full source data with a selected hyperparameter 
    \item \textit{Unsupervised learning on the target via the shadow kernel.}
    A natural label-free approach is to apply unsupervised learning directly to the target shadows by using a similarity measure between classical shadows.
We use the shadow kernel~\cite{Huang2022MBSci}
\begin{align}
    &k_{\mathrm{sh}}\!\left(S_{T_{\tgt}}(\rho^{\tgt}_j),\, S_{T_\tgt}(\rho^{\tgt}_{j'})\right)\nonumber\\
    \qquad :=& \exp(\frac{\tau}{T_{\tgt}^2}\sum_{l,l'=1}^{T_{\tgt}} \exp(\frac{\gamma}{n}\sum_{j,j'=1}^n \Tr(\sigma_{j,l}\sigma_{j',l'}))),
\end{align}
where $\sigma_{j,l}:= 3\ketbra{s_{j,l}}{s_{j,l}} - I$,
and construct the target Gram matrix
\begin{align}  
  K^{\tgt}_{jj'}
  :=&
  k_{\mathrm{sh}}\!\left(S_{T}(\rho^{\tgt}_j),\, S_{T}(\rho^{\tgt}_{j'})\right)
  \in \RR^{N_{\tgt}\times N_{\tgt}}.
  \label{eq:target-gram}
\end{align}
We then run kernel-based clustering (kernel $k$-means and spectral clustering) on $K^{\tgt}$.
The kernel hyperparameters are swept over
\begin{equation}
  \qquad \tau \in \{0.01,0.1,1.0,3.0\},
  \,
  \gamma \in \{0.01,0.1,1.0\}.
\end{equation}
Although label information is not available within the unsupervised learning, cluster indices are reordered after training using the hidden target labels to maximize the score for comparison with the UDA method.
Of course, the target labels are never used in the training or model selection.
\end{enumerate}

\subsection{Classification of quantum phase of matter}\label{subsec:qpt}

In this subsection, we consider phase classification tasks on quantum many-body datasets of 1D cluster model, 1D axial next-nearest-neighbor Ising (ANNNI) model~\cite{PhysRev.124.346}, toric codel~\cite{Kitaev2003Toric} and color code~\cite{Bombin2006Color}  systems.
Each data point is associated with the ground state of a Hamiltonian of the system.
Although we conduct the classification on the data of finite-size systems,
the task is to predict the quantum phase of the corresponding ground state defined in the thermodynamic limit.

In this task, domain adaptation is motivated by the practical discrepancy between labeled data and the data of interest at deployment.
Labeled source data are typically available in regimes where the Hamiltonian is easier to simulate or the states are easier to prepare, whereas the target regime can be harder both to prepare accurately and to label by an accurate computation.
Moreover, the target data can differ from the source data not only because the Hamiltonian parameters are distributed differently, but also because the prepared states are only approximate ground states and are further affected by hardware noise such as depolarizing errors.
Therefore, even when the target task is the same phase classification problem, a model trained only on labeled source data need not generalize well to the unlabeled target domain, making unsupervised domain adaptation a natural framework in our setting.



\begin{figure*}[t]
\centering
\begin{minipage}{0.49\textwidth}\centering
\includegraphics[width=\linewidth]{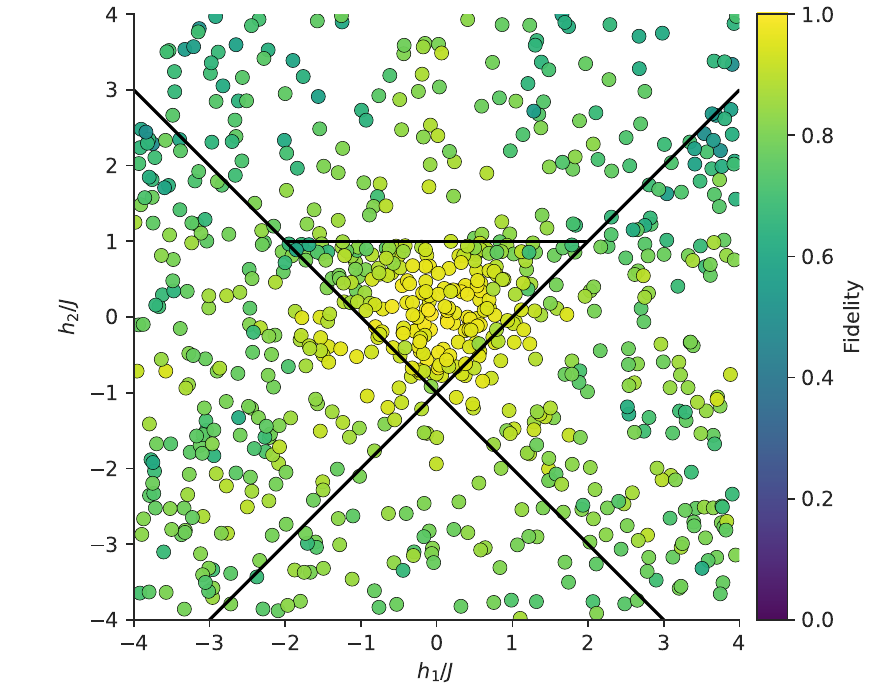}\\
\small (a) Cluster model
\end{minipage}\hfill
\begin{minipage}{0.49\textwidth}\centering
\includegraphics[width=\linewidth]{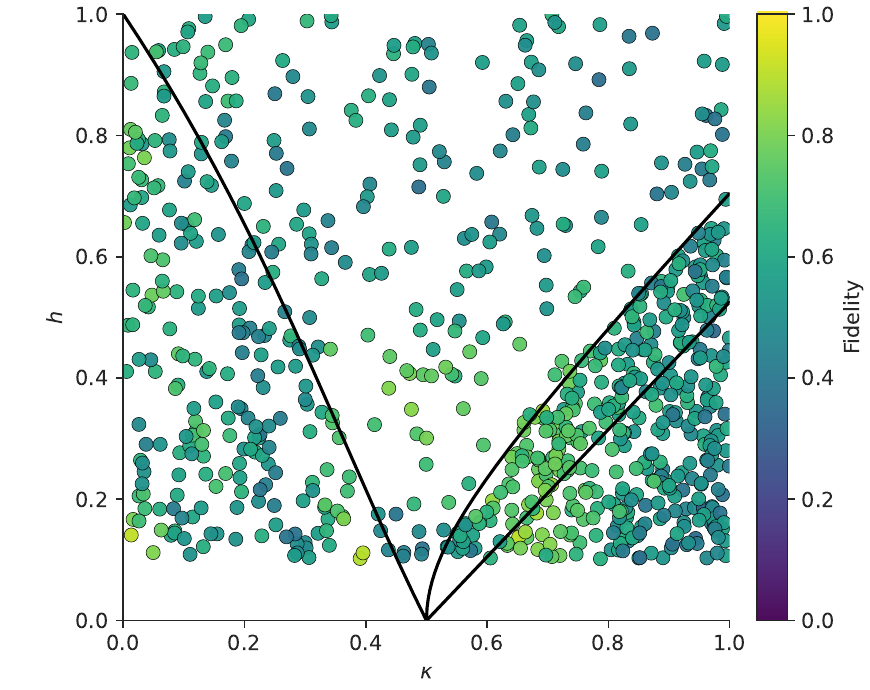}\\
\small (b) ANNNI model
\end{minipage}
\caption{
Ground-state subspace overlap of the target domains for (a) Cluster and (b) ANNNI models with QETU-based algorithmic state-preparation imperfection.
Each panel shows
$F_{\mathcal{G}}(\bm{x})=\langle \psi | \Pi_{\mathcal{G}(\bm{x})} | \psi \rangle$
at the corresponding Hamiltonian parameters.
Only target domains are shown because source states are exact in these simulations.
}
\label{fig:spinchain-fidelity}
\end{figure*}

\subsubsection{Cluster Hamiltonian}
\label{subsubsec:cluster_phase}

The Cluster Hamiltonian is defined as
\begin{equation}
  H
  = \sum_{j=1}^{n} \Bigl(
      J Z_{j}
      - h_{1} X_{j} X_{j+1}
      - h_{2} X_{j-1} Z_{j} X_{j+1}
    \Bigr),
\end{equation}
which supports multiple phases including a symmetry-protected topological (SPT), trivial, ferromagnetic, and anti-ferromagnetic phases depending on the parameters $\bm{x} = (h_1/J, h_2/J)$.
We consider a system of $n=15$ spins.
In this model, we use the input feature map $\Phi_k^1$ with $k=1$ defined in Sec.~\ref{subsec:shadow-features} as a minimal choice.

\medskip
\paragraph{Error models.}\label{para:cluster-EM}
We model algorithmic imperfections by replacing the ideal ground state with a superposition of low-energy eigenstates.


Let $\mathcal{G}(\bm{x}) \subset \mathcal{H}_n$ denote the low-energy subspace that we use as the ground-state subspace at parameter $\bm{x}$, and let $\Pi_{\mathcal{G}(\bm{x})}$ be the projector onto it.
Since all phases of this model are gapped phases, $\mathcal{G}(\bm{x})$ is taken as the span of the lowest $d_0$ eigenstates, where $d_0$ matches the degeneracy of the ground states in the thermodynamic limit.
Concretely, we use $d_0=4$ in the SPT phase, $d_0=2$ in the ferromagnetic and the anti-ferromagnetic phases, and $d_0=1$ in the trivial phase.
Then, we quantify the algorithmic imperfection in the preparation of the ground state $\ket{\psi}$ at the parameter $\bm{x}$ by its overlap with the subspace $\mathcal{G}(\bm{x})$
\[
F_{\mathcal{G}}(\bm{x})=\langle \psi | \Pi_{\mathcal{G}(\bm{x})} | \psi \rangle .
\]
For each Hamiltonian parameter $\bm{x}$, we compute the lowest $N_{\mathrm{exc}}+1$ eigenpairs with $N_{\mathrm{exc}}=40$ using sparse exact diagonalization based on the implicitly restarted Lanczos method via SciPy~\cite{Virtanen2020SciPy,Lehoucq1998ARPACK}.

Specifically, we model a two-stage imperfect state-preparation pipeline and consider two target-state constructions obtained by truncating this pipeline at different stages.
\begin{itemize}
    \item The first construction models the output of the initial preparation stage before any subsequent spectral filtering.
We therefore denote the state by $\ket{\psi_{\mathrm{raw}}(\bm{x})}$, where ``raw'' indicates the raw output of the initial state preparation routine.
Namely, we use this model to represent a generic output of approximate ground-state preparation routines that only partially reach the desired low-energy sector.
For instance, near-term ground-state preparation algorithms such as the variational quantum eigensolver, quantum imaginary time evolution, and adiabatic state preparation~\cite{Peruzzo2014,Motta2020,Albash2018} may produce such states.
In such settings, finite circuit depth, imperfect optimization, or limited evolution time can leave non-negligible excited-state contamination.
To model this scenario, the fidelity parameter $f$ is drawn uniformly from the interval $[0.2,0.4]$; then, the states $\ket{v_g}$ and $\ket{v_e}$ are uniformly sampled from $\mathcal{G}(\bm{x})$ and its orthogonal complement within the span of the computed eigenvectors, respectively.
Concretely, the raw target state is defined as
\[
\ket{\psi_{\mathrm{raw}}(\bm{x})}
=
\sqrt{f}\,\ket{v_g}
+
\sqrt{1-f}\,\ket{v_e}.
\]

    \item The second construction extends the same pipeline to the early fault-tolerant regime, where more advanced quantum algorithms become accessible.
As a representative example, we consider a spectral filtering step tailored to this setting.
Given an initial state with nonzero overlap with the ground or low-energy subspace, such methods suppress high-energy components while amplifying the desired overlap~\cite{Lin2020GroundState,Dong2022QETU}.
Accordingly, this construction represents the same imperfect initial preparation followed by an additional filtering stage, and therefore typically yields a higher overlap with $\mathcal{G}(\bm{x})$ than the raw state in the first model.

In particular, we apply the QETU-based spectral filter proposed in Ref.~\cite{Dong2022QETU} to the input state $\ket{\psi_{\mathrm{raw}}(\bm{x})}$.
We use a filter $P_{k_{\rm{deg}}}\left(\cos (\tilde{H}(\bm{x})/2)\right)$ constructed from even Chebyshev polynomials of up to degree $k_{\rm{deg}}=40$,
where $\tilde{H}(\bm{x})$ is an affine-transformed Hamiltonian whose spectrum is contained in $[\eta,\pi-\eta]$ with $\eta=0.05$.
The resulting target state is
\[
\ket{\psi_{\mathrm{QETU}}(\bm{x})}
=
\frac{
P_{k_{\rm{deg}}}\left(\cos (\tilde{H}(\bm{x})/2)\right)\ket{\psi_{\mathrm{raw}}(\bm{x})}
}{
\left\|P_{k_{\rm{deg}}}\left(\cos (\tilde{H}(\bm{x})/2)\right)\ket{\psi_{\mathrm{raw}}(\bm{x})}\right\|
}.
\]
In the generation of numerical data, we do not simulate the QETU circuit itself.
Instead, the induced spectral filter is directly applied to the computed eigenstates.
For the polynomial $P_{k_{\rm{deg}}}$, we need an estimate of the spectral gap~\cite{Dong2022QETU} and hence use a fixed gap estimate $\Delta_{\mathrm{est}}=1$ for all $\bm{x}$.
Further details of both target constructions are given in Appendix~\ref{app:target-state-construction}.
Fig.~\ref{fig:spinchain-fidelity}(a) shows the resulting target-state overlap $F_\mathcal{G}(\bm{x})$ with the subspace $\mathcal{G}(\bm{x})$.
\end{itemize}

After either algorithmic imperfection is applied, hardware-type noise is included by applying single-qubit depolarizing noise and measurement bit-flip noise with rates $p_{\mathrm{depol}}=0.1$ and $p_{\mathrm{flip}}=0.01$.
\medskip
\paragraph{Domain shift.}
In the source domain, the Hamiltonian parameters are sampled from
\begin{align*}    
    R_{\src} :=& \{(h_1/J,0)\mid -4 \le h_1/J \le 4\} \\ &\cup \{(0,h_2/J)\mid -4 \le h_2/J \le 4 \},
\end{align*}
where the source region is chosen as the union of two exactly solvable lines~\cite{Pfeuty1970TFIM,Smacchia2011ClusterIsing}.
The features are computed from exact expectation values of the clean ground states without any error, corresponding to exact simulation or to the limit $T_{\src}\rightarrow \infty$.
In fact, when the source states are classically simulable, exact computation of the expectation values can be easier than sampling.
We denote the resulting source dataset by $R_{\src}$-Exact-$\infty$.

The target-domain parameters are sampled from
\[
R_{\tgt}=\{(h_1/J,h_2/J)\mid -4 \le h_1/J, h_2/J \le 4\}\backslash R_{\src},
\]
which induces a distribution shift in the Hamiltonian parameters.
Moreover, the target states are generated by the two-stage imperfect state-preparation pipeline above.
That is, we either stop at the raw preparation stage and use $\ket{\psi_{\mathrm{raw}}(\bm{x})}$, or we further apply the QETU filter to obtain $\ket{\psi_{\mathrm{QETU}}(\bm{x})}$.
In both cases, the resulting states are subject to hardware-type noise as described in Sec.~\ref{para:cluster-EM}.
The target features are then computed from Pauli classical shadows with $T_{\tgt}=10^3$ or $10^4$ shots per data point.
The resulting target datasets are denoted by $R_{\tgt}$-Raw-$10^3$, $R_{\tgt}$-Raw-$10^4$, $R_{\tgt}$-QETU-$10^3$, and $R_{\tgt}$-QETU-$10^4$, for the four combinations considered here.



\medskip
\paragraph{Results and discussion.}
Table~\ref{table:phase}(a) shows that UDA consistently outperforms the source-only ERM in all four settings.
For Raw targets, the median macro-F1 improves from $0.664$ to $0.857$ at $T_{\tgt}=10^3$ and from $0.686$ to $0.878$ at $T_{\tgt}=10^4$.
For QETU targets, it improves from $0.678$ to $0.896$ at $T_{\tgt}=10^3$ and from $0.746$ to $0.922$ at $T_{\tgt}=10^4$.
The corresponding means show the same trend: for Raw targets, the score improves from $0.657 \pm 0.093$ to $0.855 \pm 0.030$ at $T_{\tgt}=10^3$ and from $0.693 \pm 0.085$ to $0.873 \pm 0.025$ at $T_{\tgt}=10^4$, while for QETU targets it improves from $0.676 \pm 0.107$ to $0.870 \pm 0.069$ at $T_{\tgt}=10^3$ and from $0.733 \pm 0.102$ to $0.889 \pm 0.064$ at $T_{\tgt}=10^4$ under EnsV, indicating that the improvement is visible not only in the median trial but also in the average performance over trials.
Thus, the main gain comes from adaptation rather than from merely increasing the number of target shots.
Indeed, in both the Raw and QETU settings, UDA with EnsV at $T_{\tgt}=10^3$ already exceeds the source-only baseline at $T_{\tgt}=10^4$.
At a fixed shot number, the QETU targets are also slightly easier than the Raw targets, which is consistent with the additional spectral-filtering stage improving the low-energy overlap before the hardware-type noise is applied.
It is also noteworthy that, although the setups are not directly comparable because the system sizes, dataset constructions, and evaluation metrics differ, the UDA medians at $T_{\tgt}=10^3$ are already numerically comparable to the clean supervised test accuracy reported for the cluster benchmark in Ref.~\cite{Bermejo2024QCNNSimulable}, which is $84.0\%$ for 200 training states with 4,000 shadows per state.
In this sense, adaptation on imperfect unlabeled target data already reaches the same numerical regime as recent clean-data supervised phase-classification studies.

The comparison with the shadow-kernel baselines in Table~\ref{table:phase}(b) further highlights the advantage of UDA.
The best kernel-method baseline reaches medians of only $0.652$ for Raw targets and $0.683$ for QETU targets at $T_{\tgt}=10^3$, both far below UDA at the same shot number.
Remark that the results of the shadow-kernel methods are reported only for $T_{\tgt}=10^3$.
This is because computing the shadow kernel in Eq.~\eqref{eq:target-gram} involves a double sum over the $T_{\tgt}$ snapshots, and hence the computational cost scales quadratically in $T_{\tgt}$.
Increasing the number of target shots from $10^3$ to $10^4$ therefore makes this step $10^2$ times more expensive.
Accordingly, the $T_{\tgt}=10^4$ results are omitted for these baselines.

The predicted phase diagrams in Figs.~\ref{fig:phase_cdan_10000}(a) and~\ref{fig:spinchain-phase-baselines}(a),~(c) are consistent with the above observation of the overall classification score.
UDA recovers the overall four-phase structure over most of the target region, including correct extension away from the source-supported solvable lines, whereas the source-only ERM degrade substantially once one moves off those lines.
More specifically, Fig.~\ref{fig:spinchain-fidelity}(a) shows that the target-state overlap is highest in the central region but is appreciably reduced over broad outer sectors.
Even in those regions away from the near-unit-overlap part of the target domain, UDA results in Fig.~\ref{fig:phase_cdan_10000}(a) still shows correct phase assignment over much of the parameter space.
By contrast, the source-only and shadow-kernel baselines in Fig.~\ref{fig:spinchain-phase-baselines}(a) and~\ref{fig:spinchain-phase-baselines}(c) deteriorate much more visibly there.
This indicates that the UDA model is not merely tracking the absolute overlap with $\mathcal{G}(x)$, but can still extract phase-discriminative information from imperfect shadow data once the domain shift is handled appropriately.
The remaining UDA errors are concentrated mainly near the phase boundaries and in parts of the outer sectors, which is also consistent with the standard observation in clean supervised QCNN studies that the hardest points tend to lie near phase transitions \cite{Bermejo2024QCNNSimulable}.

Regarding model selection, EnsV and InfoMax perform comparably for this benchmark, though EnsV gives the best median in all four settings, while InfoMax gives the best mean only for the QETU target with $T_{\tgt}=10^4$.
This suggests that both criteria can identify strong models in this relatively structured benchmark, with EnsV showing slightly more stable trial-to-trial behavior.

\begin{table*}[p]
\centering
\caption{
Phase-classification results measured by the macro-F1 score for the unseen target data.
Entries are mean $\pm$ standard deviation over $10$ trials with each model selection method, and the median is shown in parentheses.
Table (a) reports the results of the source-only ERM and UDA, while Table (b) reports the results of the unsupervised clustering methods based on the shadow kernel.
The dataset labels follow the common format $(\text{domain specification})\text{-}(\text{state/noise setting})\text{-}T_{\tgt}$, where the first part specifies which region of the underlying data-generating parameters is sampled, the second part specifies how the quantum data are constructed in that domain, and the last part gives the number of measurement shots per data point.
Accordingly, the labels in (a) are written as source dataset $\rightarrow$ target dataset, whereas those in (b) show only the target dataset because the shadow-kernel baselines use no source data.
For the cluster and ANNNI models, $R_{\src}$ and $R_{\tgt}$ denote the source and target Hamiltonian-parameter regions, Exact-$\infty$ means that the source features are computed from exact expectation values of clean ground states, and QETU(Raw)-$T_{\tgt}$ specifies the target-state construction together with the number of target shots per data point.
For the toric and color codes, $\mathrm{D}_1$ and $\mathrm{D}_{\le d/2}$ specify the source and target ranges of the random-circuit depth, Clean and Noisy indicate whether the stabilizer circuits are simulated without or with the injected device-inspired noise model, and the final number again denotes the number of shots per data point.
For each row in (a), the best mean and the best median among the three methods are shown in bold independently.
For each row in (b), the best mean and the best median among the six entries are shown in bold independently.
}\label{table:phase}
\textbf{(a) Source-only ERM and the UDA}\\[2pt]
\begin{tabular}{lccc}
\toprule
\multirow{2}{*}{\textbf{Task}} &
\textbf{Source only ERM} &
\multicolumn{2}{c}{\textbf{UDA}} \\
& Cross validation & EnsV & InfoMax \\
\midrule

\multirow{2}{*}{\shortstack[l]{\textbf{Cluster}\\$R_{\src}$-Exact-$\infty \rightarrow R_{\tgt}$-Raw-$10^3$}} &
$0.657 \pm 0.093$ &
$\boldsymbol{0.855 \pm 0.030}$ &
$0.844 \pm 0.029$ \\
& $(0.664)$ & $(\boldsymbol{0.857})$ & $(0.849)$ \\
\midrule

\multirow{2}{*}{\shortstack[l]{\textbf{Cluster}\\$R_{\src}$-Exact-$\infty \rightarrow R_{\tgt}$-QETU-$10^3$}} &
$0.676 \pm 0.107$ &
$\boldsymbol{0.870 \pm 0.069}$ &
$0.856 \pm 0.065$ \\
& $(0.678)$ & $(\boldsymbol{0.896})$ & $(0.870)$ \\
\midrule

\multirow{2}{*}{\shortstack[l]{\textbf{Cluster}\\$R_{\src}$-Exact-$\infty \rightarrow R_{\tgt}$-Raw-$10^4$}} &
$0.693 \pm 0.085$ &
$\boldsymbol{0.873 \pm 0.025}$ &
$0.862 \pm 0.024$ \\
& $(0.686)$ & $(\boldsymbol{0.878})$ & $(0.863)$ \\
\midrule

\multirow{2}{*}{\shortstack[l]{\textbf{Cluster}\\$R_{\src}$-Exact-$\infty \rightarrow R_{\tgt}$-QETU-$10^4$}} &
$0.733 \pm 0.102$ &
$0.889 \pm 0.064$ &
$\boldsymbol{0.905 \pm 0.028}$ \\
& $(0.746)$ & $(\boldsymbol{0.922})$ & $(0.914)$ \\
\modelrule

\multirow{2}{*}{\shortstack[l]{\textbf{ANNNI}\\$R_{\src}$-Exact-$\infty \rightarrow R_{\tgt}$-Raw-$10^3$}} &
$0.495 \pm 0.084$ &
$\boldsymbol{0.812 \pm 0.027}$ &
$0.730 \pm 0.084$ \\
& $(0.468)$ & $(\boldsymbol{0.813})$ & $(0.735)$ \\
\midrule

\multirow{2}{*}{\shortstack[l]{\textbf{ANNNI}\\$R_{\src}$-Exact-$\infty \rightarrow R_{\tgt}$-QETU-$10^3$}} &
$0.527 \pm 0.096$ &
$\boldsymbol{0.826 \pm 0.036}$ &
$0.737 \pm 0.053$ \\
& $(0.531)$ & $(\boldsymbol{0.830})$ & $(0.722)$ \\
\midrule

\multirow{2}{*}{\shortstack[l]{\textbf{ANNNI}\\$R_{\src}$-Exact-$\infty \rightarrow R_{\tgt}$-Raw-$10^4$}} &
$0.529 \pm 0.064$ &
$\boldsymbol{0.846 \pm 0.029}$ &
$0.722 \pm 0.121$ \\
& $(0.539)$ & $(\boldsymbol{0.855})$ & $(0.755)$ \\
\midrule

\multirow{2}{*}{\shortstack[l]{\textbf{ANNNI}\\$R_{\src}$-Exact-$\infty \rightarrow R_{\tgt}$-QETU-$10^4$}} &
$0.598 \pm 0.049$ &
$\boldsymbol{0.845 \pm 0.047}$ &
$0.722 \pm 0.052$ \\
& $(0.594)$ & $(\boldsymbol{0.856})$ & $(0.722)$ \\
\modelrule

\multirow{2}{*}{\shortstack[l]{\textbf{Toric code}\\$\mathrm{D}_1$-Clean-$500$ $\rightarrow$ $\mathrm{D}_{\le d/2}$-Noisy-$500$}} &
$0.779 \pm 0.011$ &
$0.834 \pm 0.028$ &
$\boldsymbol{0.886 \pm 0.019}$ \\
& $(0.784)$ & $(0.840)$ & $(\boldsymbol{0.883})$ \\
\modelrule

\multirow{2}{*}{\shortstack[l]{\textbf{Color code}\\$\mathrm{D}_1$-Clean-$500$ $\rightarrow$ $\mathrm{D}_{\le d/2}$-Noisy-$500$}} &
$0.745 \pm 0.022$ &
$0.841 \pm 0.033$ &
$\boldsymbol{0.897 \pm 0.043}$ \\
& $(0.747)$ & $(0.832)$ & $(\boldsymbol{0.889})$ \\
\bottomrule
\end{tabular}

\vspace{6pt}

\textbf{(b) Unsupervised clustering using shadow kernel}\\[2pt]
\begin{tabular}{lcccccc}
\toprule
\multirow{2}{*}{\textbf{Task}} &
\multicolumn{2}{c}{\textbf{$k$-means}} &
\multicolumn{2}{c}{\textbf{Spectral}} &
\multicolumn{2}{c}{\textbf{PCA + $k$-means}} \\
& EnsV & InfoMax & EnsV & InfoMax & EnsV & InfoMax \\
\midrule

\multirow{2}{*}{\shortstack[l]{\textbf{Cluster}\\$R_{\tgt}$-Raw-$10^3$}} &
$\boldsymbol{0.657 \pm 0.079}$&
$\boldsymbol{0.657 \pm 0.079}$&
$0.648 \pm 0.027$ &
$0.648 \pm 0.027$ &
$0.584 \pm 0.071$ &
$0.556 \pm 0.043$ \\
& $(0.628)$ & $(0.628)$ & $(\boldsymbol{0.652})$ & $(\boldsymbol{0.652})$ & $(0.583)$ & $(0.573)$ \\
\midrule

\multirow{2}{*}{\shortstack[l]{\textbf{Cluster}\\$R_{\tgt}$-QETU-$10^3$}} &
$0.596 \pm 0.067$&
$0.655 \pm 0.050$&
$\boldsymbol{0.684 \pm 0.029}$ &
$\boldsymbol{0.680 \pm 0.024}$ &
$0.619 \pm 0.046$ &
$0.592 \pm 0.055$ \\
&$(0.573)$ & $(0.677)$ & $(\boldsymbol{0.683})$ & $(\boldsymbol{0.683})$ & $(0.635)$ & $(0.611)$ \\
\modelrule

\multirow{2}{*}{\shortstack[l]{\textbf{ANNNI}\\$R_{\tgt}$-Raw-$10^3$}} &
$0.628 \pm 0.082$&
$0.671 \pm 0.070$&
$\boldsymbol{0.707 \pm 0.014}$ &
$0.705 \pm 0.014$ &
$0.583 \pm 0.039$ &
$0.566 \pm 0.034$ \\
& $(0.615)$ & $(0.694)$& $(\boldsymbol{0.706})$ & $(0.698)$ & $(0.577)$ & $(0.555)$ \\
\midrule

\multirow{2}{*}{\shortstack[l]{\textbf{ANNNI}\\$R_{\tgt}$-QETU-$10^3$}} &
$0.692 \pm 0.081$&
$0.661 \pm 0.087$&
$\boldsymbol{0.696 \pm 0.032}$ &
$0.696 \pm 0.037$ &
$0.558 \pm 0.039$ &
$0.547 \pm 0.028$ \\
& $(\boldsymbol{0.725})$ & $(0.706)$ & $(0.704)$ & $(0.703)$ & $(0.562)$ & $(0.547)$ \\
\modelrule

\multirow{2}{*}{\shortstack[l]{\textbf{Toric}\\$\mathrm{D}_{\le d/2}$-Noisy-$500$}} &
$0.441 \pm 0.176$ &
$0.488 \pm 0.195$ &
$0.537 \pm 0.033$ &
$0.526 \pm 0.022$ &
$0.493 \pm 0.104$ &
$\boldsymbol{0.762 \pm 0.014}$ \\
& $(0.342)$ & $(0.368)$ & $(0.533)$ & $(0.526)$ & $(0.472)$ & $(\boldsymbol{0.761})$ \\
\modelrule

\multirow{2}{*}{\shortstack[l]{\textbf{Color}\\$\mathrm{D}_{\le d/2}$-Noisy-$500$}} &
$0.469 \pm 0.158$ &
$0.465 \pm 0.166$ &
$0.538 \pm 0.030$ &
$0.535 \pm 0.021$ &
$0.499 \pm 0.072$ &
$\boldsymbol{0.710 \pm 0.028}$ \\
& $(0.385)$ & $(0.352)$ & $(0.535)$ & $(0.535)$ & $(0.485)$ & $(\boldsymbol{0.705})$ \\
\bottomrule
\end{tabular}
\end{table*}

\subsubsection{ANNNI model}\label{subsubsec:annni_phase}

\begin{figure*}[t]
\centering
\begin{minipage}{0.49\textwidth}\centering
\includegraphics[width=\linewidth]{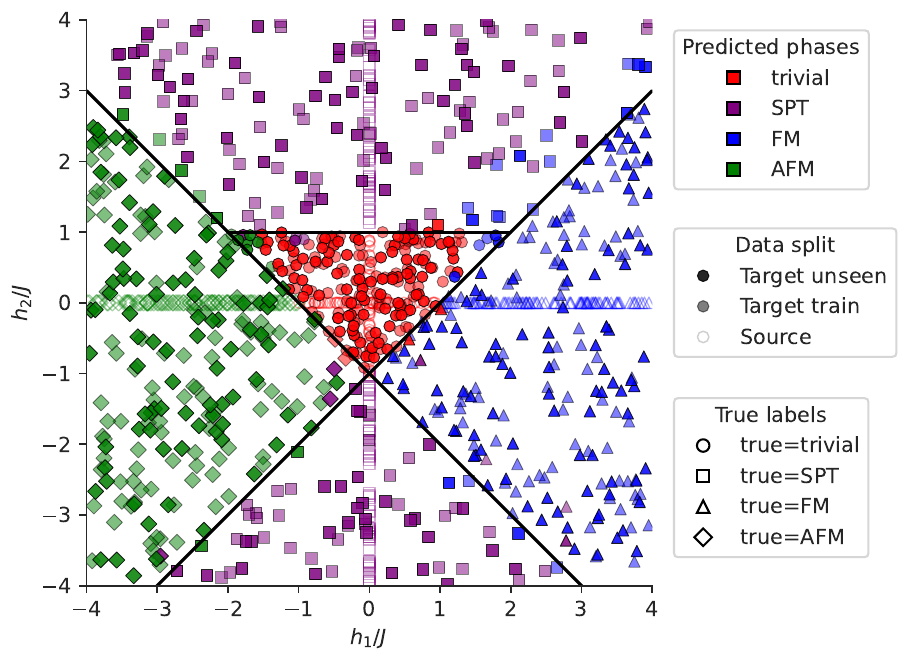}\\
\small (a) Cluster (UDA, EnsV)
\end{minipage}\hfill
\begin{minipage}{0.49\textwidth}\centering
\includegraphics[width=\linewidth]{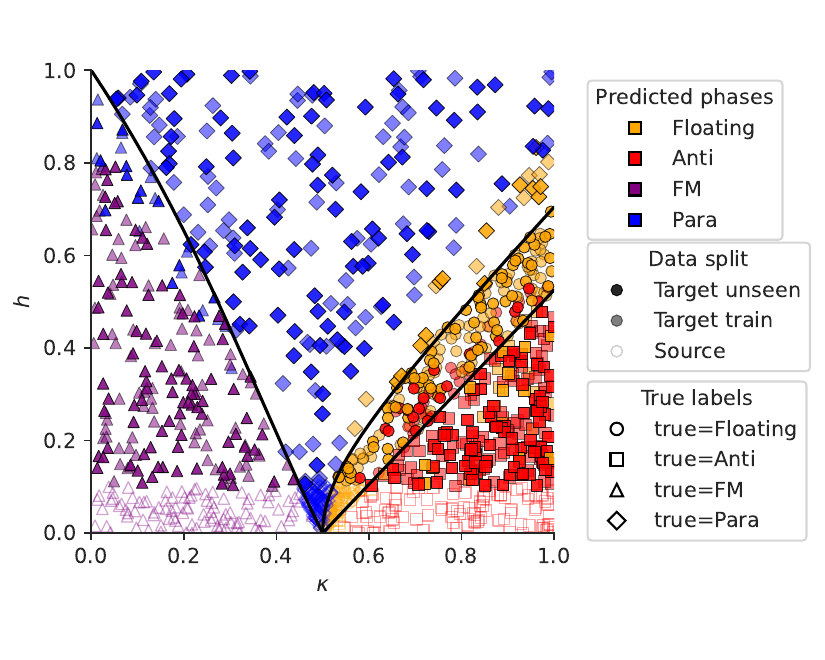}\\
\small (b) ANNNI (UDA, EnsV)
\end{minipage}
\caption{
Predicted phase diagrams for QETU-based targets obtained by UDA at $T=10^4$ shots per data point.
The displayed trial is selected as follows:
for each of the $10$ trials, select a model by EnsV using only target-train predictions, and then choose the trial whose target-unseen macro-F1 is closest to the median over the $10$ trials.
}\label{fig:phase_cdan_10000}
\label{fig:spinchain-phase-cdan}
\end{figure*}

\begin{figure*}[t]
\centering
\begin{minipage}{0.49\textwidth}\centering
\includegraphics[width=\linewidth]{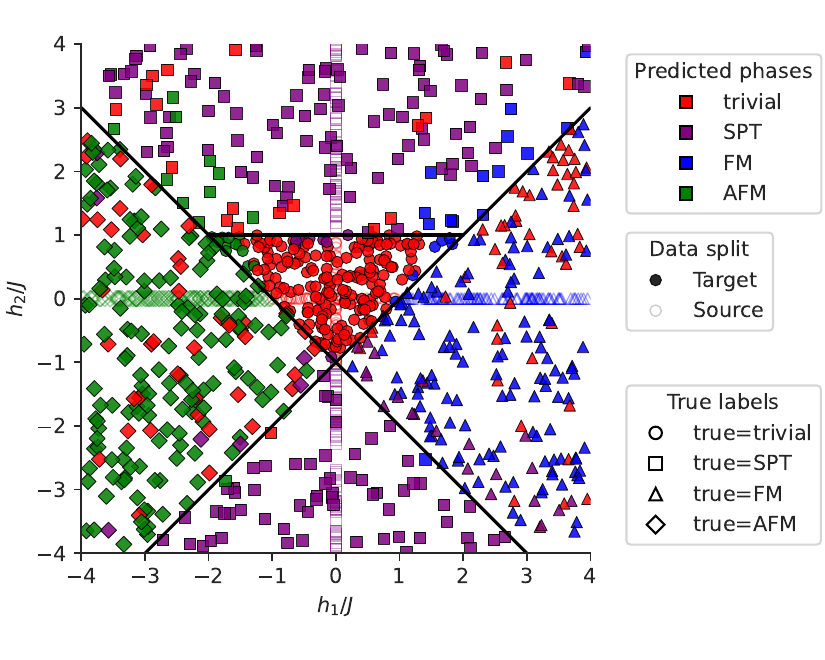}\\
\small (a) Cluster (source-only ERM, $10^4$ shots per data point)
\end{minipage}\hfill
\begin{minipage}{0.49\textwidth}\centering
\includegraphics[width=\linewidth]{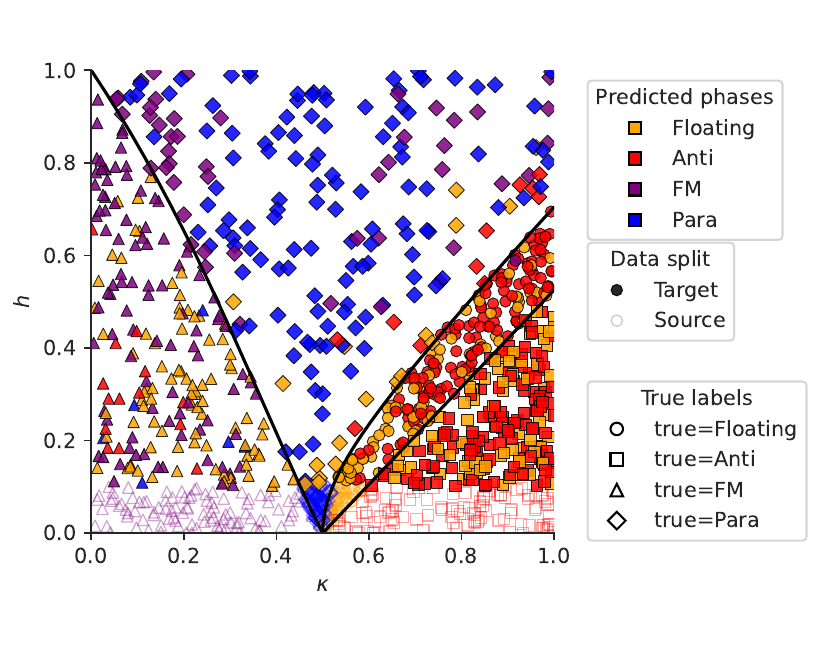}\\
\small (b) ANNNI (source-only ERM, $10^4$ shots per data point)
\end{minipage}

\vspace{0.8em}
\begin{minipage}{0.49\textwidth}\centering
\includegraphics[width=\linewidth]{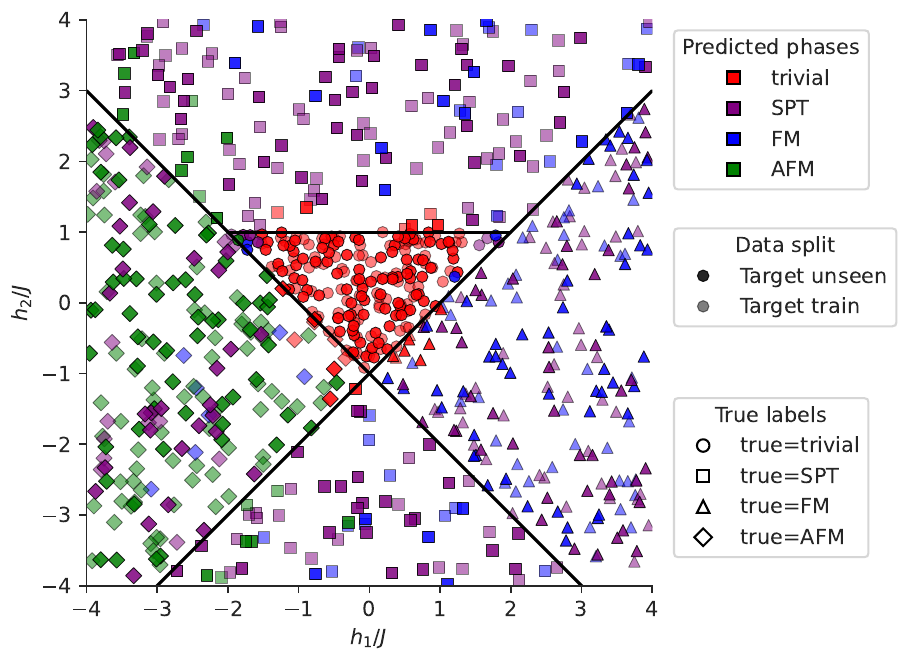}\\
\small (c) Cluster (shadow-kernel spectral, EnsV, $10^3$ shots per data point)
\end{minipage}\hfill
\begin{minipage}{0.49\textwidth}\centering
\includegraphics[width=\linewidth]{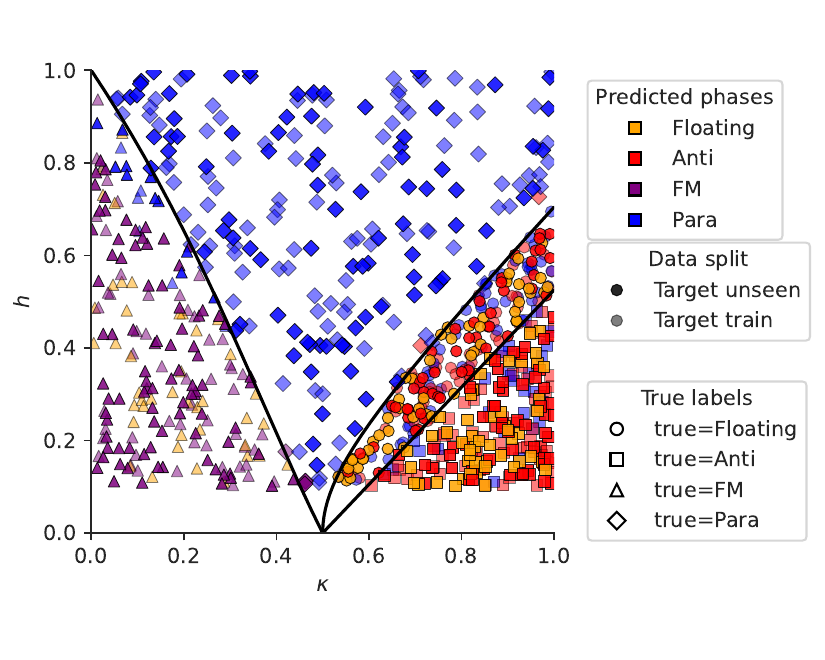}\\
\small (d) ANNNI (shadow-kernel $k$-means, EnsV, $10^3$ shots per data point)
\end{minipage}

\caption{
Predicted phase diagrams of baseline methods for QETU targets.
Top row: source-only ERM with $10^4$ shots per data point.
In each trial, the model is selected by source-domain cross-validation, and the displayed trial is the one whose target-unseen macro-F1 is closest to the median over the $10$ trials.
Bottom row: unsupervised clustering baselines based on the shadow kernel with $10^3$ shots per data point.
For each benchmark, we show the shadow-kernel clustering method that attains the best median target-unseen macro-F1 among the unsupervised shadow-kernel baselines.
In each trial, hyperparameters are selected by the corresponding unlabeled criterion using only target-train data, and the displayed trial is the one whose target-unseen macro-F1 is closest to the median over the $10$ trials.
}
\label{fig:spinchain-phase-baselines}
\end{figure*}

The ANNNI model is defined as
\begin{equation}
  H
  = - J_{1} \sum_{j=1}^{n-1} X_{j} X_{j+1}
    - J_{2} \sum_{j=1}^{n-2} X_{j} X_{j+2}
    - B \sum_{j=1}^{n} Z_{j},
\end{equation}
which has ferromagnetic phase, antiphase, floating phase, and paramagnetic phase
depending on the parameters $\bm{x}= (\kappa, h)$
with $\kappa=-J_{2}/J_{1}$ and $h = B/J_{1}$ \cite{PhysRev.124.346}.
Especially, floating phase is a gapless phase with power-law-decaying correlations and is not guaranteed to be efficiently classified by classical shadows \cite{Huang2022MBSci}.
We also consider $n=15$ spins.
In this model, we use the input feature map $\Phi_k^1$ with $k=4$ defined in Sec.~\ref{subsec:shadow-features} to capture correlations of distant sites.

\medskip
\paragraph{Error models.}\label{para:annni-EM}
We use the same error model of the algorithmic imperfection and the hardware-type noise as that for the cluster model specified in Sec.~\ref{para:cluster-EM}.
For the gapped phases, as in the cluster model, $\mathcal{G}(\bm{x})$ is taken as the span of the lowest $d_0$ eigenstates, where $d_0$ is chosen to match the ground-state degeneracy of that phase in the thermodynamic limit.
Accordingly, we use $d_0=2$ in the ferromagnetic phase, $d_0=4$ in the anti-phase, and $d_0=1$ in the paramagnetic phase.
For the gapless floating phase, $\mathcal{G}(\bm{x})$ is taken as the span of eigenstates in the band
\(
E \le E_0 + 3(E_1-E_0),
\)
where $E_0$ and $E_1$ are the smallest and second-smallest eigenvalues at $\bm{x}$.
As for an estimate $\Delta_{\mathrm{est}}$ of the spectral gap for the construction of QETU-based spectral filter, we use $0.01$ in this model.
Fig.~\ref{fig:spinchain-fidelity}~(b) shows the resulting target state overlap $F_\mathcal{G}(\bm{x})$ with the subspace $\mathcal{G}(\bm{x})$.
The remaining parameters are the same as those for the cluster model.

\medskip
\paragraph{Domain shift.}
In the source domain, the Hamiltonian parameters are sampled from
\[
R_{\src}:=\{(\kappa,h)\mid 0\le \kappa \le 1,\ 0\le h\le 0.1\},
\]
and the features are computed from exact expectation values of the clean ground states without any error, as in the cluster-model setting above.
We denote the resulting source dataset by $R_{\src}$-Exact-$\infty$.
This source region is chosen as a relatively simple low-field regime, while the target domain should cover the broader parameter region.

The target-domain parameters are sampled from
\[
R_{\tgt}:=\{(\kappa,h)\mid 0\le \kappa \le 1,\ 0\le h\le 1\}\setminus R_{\src},
\]
which induces a distribution shift in the Hamiltonian parameters.
Moreover, the target states are generated by the same two-stage imperfect state-preparation pipeline as in the cluster-model setting, with the ANNNI-specific choice of $\mathcal{G}(\bm{x})$ and $\Delta_{\mathrm{est}}$ specified in Sec.~\ref{para:annni-EM}.
That is, we either stop at the raw preparation stage and use $\ket{\psi_{\mathrm{raw}}(\bm{x})}$, or we further apply the QETU filter to obtain $\ket{\psi_{\mathrm{QETU}}(\bm{x})}$.
In both cases, the resulting states are further subject to hardware-type noise.
The target features are then computed from Pauli classical shadows with $T_{\tgt}=10^3$ or $10^4$ shots per data point.
The resulting target datasets are denoted by $R_{\tgt}$-Raw-$10^3$, $R_{\tgt}$-Raw-$10^4$, $R_{\tgt}$-QETU-$10^3$, and $R_{\tgt}$-QETU-$10^4$.


\medskip
\paragraph{Results and discussion.}
The ANNNI benchmark is clearly more challenging than the Cluster benchmark, and Table~\ref{table:phase}(a) shows a correspondingly larger gap between UDA and the baselines.
For Raw targets, the median macro-F1 improves from $0.468$ to $0.813$ at $T_{\tgt}=10^3$ and from $0.539$ to $0.855$ at $T_{\tgt}=10^4$.
For QETU targets, it improves from $0.531$ to $0.830$ at $T_{\tgt}=10^3$ and from $0.594$ to $0.856$ at $T_{\tgt}=10^4$.
The corresponding means also improve substantially: for Raw targets, the score improves from $0.495 \pm 0.084$ to $0.812 \pm 0.027$ at $T_{\tgt}=10^3$ and from $0.529 \pm 0.064$ to $0.846 \pm 0.029$ at $T_{\tgt}=10^4$, while for QETU targets it improves from $0.527 \pm 0.096$ to $0.826 \pm 0.036$ at $T_{\tgt}=10^3$ and from $0.598 \pm 0.049$ to $0.845 \pm 0.047$ at $T_{\tgt}=10^4$ under EnsV, showing that the advantage of UDA is reflected not only in the median but also in the average score over trials.
Hence, as in the cluster model, UDA already achieves strong performance with $10^3$ target shots, and increasing the shot number to $10^4$ yields a further improvement.

The gain from $10^3$ to $10^4$ is especially visible in the Raw setting, while the QETU setting is already close to saturation at $10^3$, which suggests that the remaining difficulty is dominated more by the domain shift and the phase geometry than by measurement noise alone, once the target states are partially improved by filtering.
By contrast, the source-only ERM remains much less accurate in every setting, showing that labeled data from the source region are informative but insufficient to bridge the substantial shift to the imperfect target domain.
It is again noteworthy that, although the setups are not directly comparable, the UDA medians at both $T_{\tgt}=10^3$ and $T_{\tgt}=10^4$ are in the same numerical range as the clean supervised test accuracies reported for the ANNNI benchmark in Ref.~\cite{Bermejo2024QCNNSimulable}; namely, $85.8\%$ with 200 training states and $80.2\%$ with 20 training states with 4,000 and 2,500 shadows per state.
Thus, even under domain shift and imperfect target-state generation, UDA reaches a performance level comparable in magnitude to recent clean supervised phase-classification studies.

The target-only shadow-kernel baselines are again clearly inferior.
For the Raw target at $T_{\tgt}=10^3$, the best kernel baseline attains a median of $0.706$, and for the QETU target the best median is $0.725$.
These values remain well below CDAN with EnsV, whose medians are $0.813$ and $0.830$, respectively.
Thus, although the target data contains some cluster structure, unlabeled kernel clustering alone is not sufficient in this imperfect and distribution-shifted regime.
As in the cluster-model benchmark, we report the shadow-kernel methods only for $T_{\tgt}=10^3$ because of the quadratic scaling in $T_{\tgt}$.

The predicted phase diagrams in Figs.~\ref{fig:phase_cdan_10000}(b) and~\ref{fig:spinchain-phase-baselines}(b),~(d) are consistent with the above observations.
UDA recovers the overall ANNNI phase structure reasonably well, including a recognizable floating-phase region.
This is nontrivial because the floating phase is gapless and has long-range correlations, making classification from finite-shot classical shadows more challenging than in gapped phases.

Fig.~\ref{fig:spinchain-phase-baselines}(d) illustrates that the shadow-kernel $k$-means result for the QETU target reflects this limitation through substantially stronger confusion among the floating, antiphase, and ferromagnetic regions than in UDA.
More specifically, Fig.~\ref{fig:spinchain-fidelity}(b) shows that the target-state overlap is far from uniform and is visibly reduced over broad parts of the target region, including large areas above and around the floating-phase wedge.
Nevertheless, Fig.~\ref{fig:phase_cdan_10000}(b) still recovers the wedge-shaped floating region and the surrounding ordered phases over most of the domain.
By contrast, the source-only and shadow-kernel baselines in Fig.~\ref{fig:spinchain-phase-baselines}(b) and~\ref{fig:spinchain-phase-baselines}(d) show much stronger confusion in those same parts of the parameter space.
Thus, again, good target performance is obtained not only where the overlap is relatively high, but also in substantial regions where the target states remain noticeably imperfect.
Most remaining UDA errors are concentrated near the phase boundaries, which is again consistent with the clean supervised QCNN literature and with the intrinsic difficulty of phase recognition near critical regions \cite{Bermejo2024QCNNSimulable}.

The difference between EnsV and InfoMax is much more pronounced here than in the cluster-model benchmark.
Across all four ANNNI settings in Table~\ref{table:phase}, EnsV gives uniformly higher target-unseen mean and median macro-F1 than InfoMax.
This suggests that, for this benchmark, robustness of model selection is more important than selecting a possibly better but less stable individual candidate.
One possible interpretation is that EnsV and InfoMax emphasize different tradeoffs.
EnsV uses ensemble-averaged predictions across candidate models, which can make the selection more robust to unstable candidates.
In contrast, because InfoMax evaluates candidates model by model, it can be more sensitive to mismatch between the unlabeled selection score and the true target-unseen performance, although it can in principle recover a peak-performing model.
The present ANNNI results suggest that the candidate models vary enough in quality that the robustness of EnsV is more beneficial than the peak-model selection capability of InfoMax in this benchmark.

\subsubsection{Toric code and color code datasets}
\label{subsubsec:code_phase}

The toric code model~\cite{Kitaev2003Toric} is defined as
\begin{equation}
H = - \sum_{s} A_{s} - \sum_{p} B_{p},
\end{equation}
with $A_{s} = \prod_{j \in s} X_{j}$ and $B_{p} = \prod_{j \in p} Z_{j}$.
Here $s$ and $p$ denote stars and plaquettes of a square lattice, respectively.
The color code model~\cite{Bombin2006Color} is defined as
\begin{equation}
H = - \sum_{f} \left( S^{X}_{f} + S^{Z}_{f} \right),
\end{equation}
with $S^{X}_{f} = \prod_{j \in f} X_{j}$ and $S^{Z}_{f} = \prod_{j \in f} Z_{j}$.
Here $f$ denotes the faces of a trivalent, three-colorable lattice.

To generate datasets for these models, we follow a procedure similar to Ref.~\cite{Huang2022MBSci}, where topological and trivial phases are realized by applying geometrically local random quantum circuits to distinct reference states.
In the topological phase, data points are obtained by preparing ground states of the corresponding Hamiltonians and subsequently applying a low-depth geometrically local random quantum circuit acting on neighboring qubits of the underlying lattice.
For the trivial phase, we instead start from a product state and apply the same class of geometrically local random quantum circuits.
To enable large-scale numerical simulations, with 98 qubits (code distance $d=7$) for the toric code and 91 qubits (code distance $d=11$) for the color code, both code states are simulated using a stabilizer simulator based on the Gottesman--Knill theorem.
The data are serialized into one-dimensional configurations following the qubit ordering of qecsim~\cite{qecsim} and used as input to a one-dimensional CNN.
In these models, we use the input feature map $\Phi_k^1$ with $k=1$ defined in Sec.~\ref{subsec:shadow-features} based on the above one-dimensional mapping.

\medskip
\paragraph{Error models.}
Within the procedure outlined above, algorithmic state-preparation imperfections based on low-energy superpositions are not incorporated because they do not preserve stabilizer structure.
Instead, we emulate realistic hardware noise by executing the stabilizer circuits on a fake backend that reproduces device characteristics of an IBM Quantum processor, \texttt{fake\_kawasaki}~\cite{IBMFakeKawasaki}, including qubit connectivity, available gate sets, and calibration-derived noise models.
In the target domain, we approximate the above device noise as a mixture of Pauli channels and inject this noise into the simulated circuits, whereas the source domain uses clean stabilizer circuits without this additional noise.

\medskip
\paragraph{Domain shift.}
As a shift in the distribution of the underlying ground states, we vary the depth of the geometrically local random circuits applied to the states.
In the source domain, the circuit depth is fixed to one layer, while the depth is sampled at random up to at most one half of the code distance in the target domain, resulting in the source and target data drawn from different regions of the ground states.

Following the convention used for the cluster and ANNNI datasets, we denote the source dataset by $\mathrm{D}_1$-Clean-500 and the target dataset by $\mathrm{D}_{\le d/2}$-Noisy-500.
Here, $\mathrm{D}_1$ indicates a fixed one layer for the geometrically local random-circuit, while $\mathrm{D}_{\le d/2}$ means that the depth is sampled up to one half of the code distance~$d$.
``Clean" refers to stabilizer simulations without the additional device-noise model, and ``Noisy" includes the device-inspired Pauli-channel noise.
Also, the final number denotes the number of measurement shots per data point.

\medskip
\paragraph{Results and discussion.}
Table~\ref{table:phase}(a) shows that UDA consistently outperforms the source-only ERM for both codes.
For the toric code, the median macro-F1 improves from $0.784$ to $0.840$ with EnsV and to $0.883$ with InfoMax, and the corresponding means are $0.779 \pm 0.011$, $0.834 \pm 0.028$, and $0.886 \pm 0.019$, respectively.
For the color code, the median improves from $0.747$ to $0.832$ and $0.889$, with corresponding means $0.745 \pm 0.022$, $0.841 \pm 0.033$, and $0.897 \pm 0.043$.
Thus, the advantage of UDA is visible not only in the median but also in the average performance over trials, with InfoMax achieving the best overall results for both models.

This improvement is obtained with only 500 shots per data point, which is substantially fewer than in the cluster and ANNNI benchmarks. 
A plausible interpretation is that, compared with those benchmarks, the present setting is affected by a milder effective error and domain shift, since it does not include additional algorithmic state-preparation errors. 
It is also notable that the improvement is observed even for the large system sizes, reaching 98 qubits for the toric code and 91 qubits for the color code. 
These results indicate that the proposed approach remains effective for large-scale systems subject to hardware noise and distribution shifts of the underlying ground states.
Moreover, unlike the previous one-dimensional tasks, these are intrinsically two-dimensional models. 
Nevertheless, the proposed pipeline remains effective even when the data are serialized and processed by a one-dimensional CNN, indicating that, for this task, robust phase classification does not appear to rely on an architecture that is explicitly tailored to the underlying two-dimensional lattice geometry.

The comparison with the target-only unsupervised clustering methods using shadow-kernel in Table~\ref{table:phase}(b) further highlights the advantage of UDA. 
For the toric code, the best kernel-method baseline attains a median of $0.761$, while the corresponding UDA result reaches $0.883$. 
For the color code, the best kernel baseline attains a median of $0.705$, again well below the UDA result of $0.889$. 
Hence, although the target data contain enough structure for target-only unsupervised learning to achieve nontrivial performance, direct clustering on the target shadows remains clearly inferior to adaptation with labeled source data.

For both the toric and color codes, InfoMax gives the best mean and median scores among the UDA models, whereas EnsV remains weaker though still substantially better than the baselines. 
This suggests that, in contrast to the ANNNI benchmark where EnsV was more reliable, the milder effective error and domain shift, as described in the earlier interpretation, can make InfoMax’s confidence-seeking tendency beneficial rather than misleading.

\begin{table*}[t]
\centering
\caption{
Entanglement-classification results measured by the macro-F1 score for the unseen target data.
Same conventions as in Table~\ref{table:phase}, except that the dataset labels are written as follows:
for multipartite separable(Sep)/entanglement(Ent) classification tasks,
(system size, partition size, partition structure)-(noise setting);
for GHZ/W classes classification tasks,
(system size, state-generation procedure)-(noise setting).
In all tasks, the numbers of source and target shots per data point are fixed to
$T_{\src}=T_{\tgt}=500$.
}\label{table:entangle}

\textbf{(a) Source-only ERM and the UDA}\\[2pt]
\begin{tabular}{lccc}
\toprule
\multirow{2}{*}{\textbf{Task}} &
\textbf{Source only ERM} &
\multicolumn{2}{c}{\textbf{UDA}} \\
 & Cross Validation & EnsV & InfoMax \\
\midrule

\textbf{Multipartite Sep/Ent} &
$0.796 \pm 0.180$ &
$0.667 \pm 0.278$ &
$\boldsymbol{0.903 \pm 0.040}$ \\
(8,3,fix)-Clean$\rightarrow$(16,3,random)-Noisy & 
$(0.844)$ & $(0.852)$ & $(\boldsymbol{0.898})$ \\
\midrule

\textbf{Multipartite Sep/Ent} &
$0.387 \pm 0.024$ &
$0.924 \pm 0.031$ &
$\boldsymbol{0.949 \pm 0.016}$ \\
(16,6,fix)-Clean$\rightarrow$(16,3,random)-Noisy & 
$(0.382)$ & $(0.930)$ & $(\boldsymbol{0.940})$ \\
\midrule

\textbf{GHZ/W classes} &
$0.446 \pm 0.088$ &
$0.886 \pm 0.021$ &
$\boldsymbol{0.904 \pm 0.030}$ \\
(8,slocc)-Clean$\rightarrow$(16,slocc)-Noisy & 
$(0.453)$ & $(0.886)$ & $(\boldsymbol{0.917})$ \\
\midrule
\textbf{GHZ/W classes} &
$0.924 \pm 0.026$ &
$\boldsymbol{0.937 \pm 0.023}$ &
$0.906 \pm 0.080$ \\
(16,pauli)-Clean$\rightarrow$(16,slocc)-Noisy & 
$(0.925)$ & $(\boldsymbol{0.942})$ & $(0.930)$ \\
\bottomrule
\end{tabular}

\vspace{6pt}

\textbf{(b) Unsupervised clustering using shadow kernel}\\[2pt]
\begin{tabular}{lcccccc}
\toprule
\multirow{2}{*}{\textbf{Task}} &
\multicolumn{2}{c}{\textbf{$k$-means}} &
\multicolumn{2}{c}{\textbf{Spectral}} &
\multicolumn{2}{c}{\textbf{PCA + $k$-means}} \\
 & EnsV & InfoMax & EnsV & InfoMax & EnsV & InfoMax \\
\midrule

\textbf{Multipartite Sep/Ent} &
$0.499 \pm 0.203$ &
$0.522 \pm 0.196$ &
$0.526 \pm 0.023$ &
$0.524 \pm 0.024$ &
$0.577 \pm 0.030$ &
$\boldsymbol{0.646 \pm 0.115}$ \\
(16,3,random)-Noisy &
$(0.349)$ & $(0.463)$ & $(0.521)$ & $(0.518)$ & $(0.579)$ & $(\boldsymbol{0.584})$ \\
\midrule

\textbf{GHZ/W classes} &
$0.557 \pm 0.286$ &
$0.474 \pm 0.251$ &
$0.538 \pm 0.039$ &
$0.546 \pm 0.041$ &
$0.692 \pm 0.146$ &
$\boldsymbol{0.940 \pm 0.010}$ \\
(16,slocc)-Noisy &
$(0.365)$ & $(0.359)$ & $(0.534)$ & $(0.531)$ & $(0.634)$ & $(\boldsymbol{0.942})$ \\
\bottomrule
\end{tabular}

\end{table*}


\subsection{Classification of entangled states}

As a complementary task, we consider multi-partite entanglement classification of multi-qubit systems.

The classification and certification of entanglement classes are central problems in quantum information science.
For example, detecting multipartite entanglement enables the benchmarking of a device’s capability to generate the intended quantum states, while the controlled production and verification of genuine multipartite entanglement is widely regarded as a key indicator of functional performance~\cite{friis2019entanglement,friis2018observation}. 
Furthermore, identifying entanglement structures in thermal or ground states offers insight into the classical simulability of many-body systems~\cite{friis2019entanglement}.

Despite its importance, quantifying and characterizing multipartite entanglement are challenging tasks.
Full state tomography can provide detailed information about generated states and their entanglement structure, but its resource requirements scale unfavorably with system size, rendering the approach impractical for larger systems.
In addition, various entanglement measures and witnesses have been developed for classification and certification; however, these measures are not always sufficient to distinguish different forms of multipartite entanglement and can be inefficient to evaluate in practice~\cite{Cho2024QuantumExperimentalData,Schatzki2021NTangled}


Machine learning, by contrast, provides a flexible framework for extracting relevant features directly from quantum states and has been shown to outperform the direct use of such quantities in certain settings~\cite{Luo2023GME,Vintskevich2023entanglement,Koutn2023entanglement,Cho2024QuantumExperimentalData,Huang2025entanglement}. 
This motivates us to examine our UDA approach in this task.
Specifically, we consider two discrimination tasks: (1) distinguishing between multipartite separable and genuinely entangled states, and (2) distinguishing between GHZ- and W-type entanglement under variations in system size, state-generation procedures, and subsystem partitions. Classical-shadow features are computed from randomized single-qubit Pauli measurements, with the numbers of source and target shots per data point fixed to $T_{\src} = T_{\tgt} = 500$. For both tasks, we use the input feature map $\Phi_k^1$ with $k=3$ defined in Sec.~\ref{subsec:shadow-features}.
The same CDAN-based unsupervised domain adaptation pipeline is subsequently applied.

\subsubsection{Multipartite separable/entangled states classification.}
We classify quantum states according to whether they are multipartite separable or multipartite entangled.  

\medskip
\paragraph{Error models.}
In order to account for hardware noise inherent in practical settings such as device benchmarking and many-body experiments, we introduce noise into the target domain.
As discussed below, the target domain is a more experimentally demanding regime, involving larger systems and partitions, and more complex partition structures, and is therefore more susceptible to such noise.
Specifically, we apply single-qubit depolarizing noise and measurement bit-flip noise to the target domain, with rates $p_{\mathrm{depol}}=0.1$ and $p_{\mathrm{flip}}=0.01$.

\medskip
\paragraph{Domain shift.}
Since classifying multipartite entanglement states in larger systems and under nontrivial partitions is important in practical settings~\cite{Fuchs2025entanglement,Zhou2019entanglement,Zander2024entanglement}, we examine two domain shifts in addition to the above noise in the target domain.

The first domain shift arises from a variation in system size and partition structure.
The source domain contains 8-qubit tripartite separable states in
which the subsystems are defined by a fixed partition and each subsystem is a Haar-random
pure state, together with 8-qubit Haar-random tripartite entangled states.  
The target domain contains 16-qubit tripartite separable states in which the partition is
sampled uniformly at random for each state, together with 16-qubit
tripartite entangled states. 

The second domain shift arises from a variation in partition size and partition structure.
The source domain consists of 16-qubit six-partite separable states
with a fixed partition and 16-qubit six-partite entangled states, while the target
domain is identical to that of the first setting.

To simplify the dataset labels used below, we write each dataset in this task by
(system size, partition size, partition structure)-(noise setting).
For example, (16,6,fix)-Clean denotes the dataset of 16-qubit states with a fixed six-part partition and no injected noise, while (16,3,random)-Noisy denotes the dataset of 16-qubit tripartite states with a randomly sampled partition under the noise model.

\paragraph{Results and discussion.}
The two multipartite separability benchmarks in Table~\ref{table:entangle} illustrate two qualitatively different transfer regimes.
In the first setting, $(8,3,\mathrm{fix})$-Clean$\rightarrow(16,3,\mathrm{random})$-Noisy, the source-only ERM already attains a reasonably strong median of $0.844$, indicating that the basic distinction between tripartite separable and multipartite entangled states transfers to some extent across the increase in system size.
However, its mean and standard deviation, $0.796 \pm 0.180$, show that this transfer is highly unstable across trials.
UDA with InfoMax improves both the typical and average performance to $0.903 \pm 0.040$ with median $0.898$, whereas UDA with EnsV reaches a similar median, $0.852$, but a much worse mean and a very large variance, $0.667 \pm 0.278$.
Thus, in this setting the main benefit of UDA is not only a higher median score, but also a marked stabilization of transfer when combined with an appropriate target-label-free model-selection criterion.

In the second setting, $(16,6,\mathrm{fix})$-Clean$\rightarrow(16,3,\mathrm{random})$-Noisy, the contrast is even sharper.
Here the source-only ERM collapses to $0.387 \pm 0.024$ with median $0.382$, while UDA improves the score to $0.924 \pm 0.031$ with median $0.930$ for EnsV and to $0.949 \pm 0.016$ with median $0.940$ for InfoMax.
This is the clearest separability benchmark in which non-adaptive transfer fails systematically.
A natural interpretation is that training on six-partite separability does not transfer well to tripartite separability even at fixed system size, because the geometry of the separable class changes substantially when the partition size and partition structure are altered.
In other words, the issue is not merely additional noise, but a substantial change in the effective class structure seen by the classifier.
Once unlabeled target data are incorporated, however, CDAN can adapt the representation and recover a high-accuracy labeled predictor.

The comparison with the target-only unsupervised baselines in Table~\ref{table:entangle}(b) further clarifies the role of source supervision.
For the common target dataset $(16,3,\mathrm{random})$-Noisy, the best shadow-kernel baseline attains only $0.646 \pm 0.115$ with median $0.584$.
Thus, the target-domain structure alone is not sufficient for reliable target classification in this task.
The large gap between these scores and those of UDA shows that labeled source information remains essential, but it must be transferred in an adapted manner rather than by direct source-only training.

The two settings also highlight the importance of target-label-free model selection in UDA for quantum data.
The comparison between EnsV and InfoMax suggests that, in this benchmark, the main limitation of EnsV is its tendency to align with the ensemble-averaged predictions across candidate models rather than to identify the best individual model.
Consequently, when strong individual candidates exist but the ensemble average is not itself strong, EnsV can miss the peak-performing model.
InfoMax, by contrast, evaluates candidates model by model, so it can recover such a model when its unlabeled score aligns well with target-unseen performance, although this can come at the price of greater instability.
In the present separable/entangled states classification benchmarks, the results suggest that strong individual candidate models exist and are captured more effectively by InfoMax than by EnsV, especially in the more difficult transfer settings.

\subsubsection{GHZ/W entanglement classes.}

The GHZ class and W class are defined as the sets of quantum states reachable from the
canonical $n$-qubit GHZ or W state through stochastic local operations and classical communication (SLOCC)~\cite{Dur2000entanglementSLOCC,Miyake2003entanglementSLOCC}.

\medskip
\paragraph{Error models.}
As in the multipartite separable/entangled classification tasks, we introduce hardware noise into the target domain. 
As discussed below, the experimental difficulty arises from larger systems and broader state-generation families, making the target domain more susceptible to hardware noise.
Specifically, we employ the same noise model as in the previous tasks.

\medskip
\paragraph{Domain shift.}
Motivated by the extensive studies on the classification of GHZ and W classes for multiqubit systems and families of SLOCC-equivalent states~\cite{Dur2000entanglementSLOCC,Miyake2003entanglementSLOCC,Chen2006entanglement,Vintskevich2023entanglement}, we consider two domain shifts as well as the above noise in the target domain.

The first domain shift arises from a variation in system size.
The source domain consists of 8-qubit GHZ and W class states generated by random SLOCC,
and the target domain consists of 16-qubit GHZ/W class states also generated by random SLOCC.

The second domain shift arises from a variation in state-generation procedures, so that the source and target data are effectively sampled from different parameter regions.
The source domain consists of 16-qubit GHZ and W class states created by applying Pauli operators sampled uniformly at random, and the target domain
again consists of 16-qubit GHZ and W class states generated using random SLOCC.

We write each dataset in this task by
(system size, state-generation procedure)-(noise setting).
For example, (16,pauli)-Clean denotes the dataset of 16-qubit clean states in GHZ/W classes generated by random Pauli operations, while (16,slocc)-Noisy denotes the dataset of $16$-qubit noisy states in GHZ/W classes generated in random SLOCC.

\medskip
\paragraph{Results and discussion.}
The results are summarized in Table~\ref{table:entangle}. 
In the first setting, UDA substantially outperforms the source-only ERM, indicating that domain adaptation is beneficial when the domain shift significantly degrades transfer to the target domain.
In the second setting, however, the source-only ERM already performs strongly, and UDA with the InfoMax does not improve on it.
This suggests that domain adaptation does not necessarily lead to improvements when the source model already transfers well to the target domain.
For both settings, the unsupervised baseline achieves higher scores than UDA, showing that clustering directly on target-domain data can sometimes capture the class structure more effectively, although it does not provide labeled predictors.
This suggests a potential future direction to incorporate methods that combines target-domain clustering with domain adaptation \cite{Pan2011TCA,Long2015TKL,tuia2016kernel}, where clustering captures the target-domain class structure while the adapted model provides a labeled predictor, leading to further performance improvements in such tasks.

\section{Conclusion}
\label{sec:conclusion}
Realistic learning problems with quantum data rarely provide fully labeled, in-distribution, and perfectly prepared samples. This challenge is especially relevant in many-body settings, where useful quantum data may be accessible even when exact classical simulation or labeling is costly, and where classical shadows provide a flexible and sample-efficient interface between quantum experiments and classical learning \cite{Huang2021PowerOfData,Huang2020shadow,Cho2024QuantumExperimentalData}. In this work, we studied this setting from the viewpoint of unsupervised domain adaptation and proposed a concrete pipeline that combines classical-shadow-based feature extraction, a CDAN-based adaptation model, and label-free model selection. The resulting approach learns predictors for unlabeled and imperfect target domains by leveraging labeled data from related but shifted source domains together with unlabeled target data.

Across the phase-classification and entanglement-classification benchmarks, the proposed pipeline substantially improves over the source-only ERM baseline and, in many cases, also over target-only unsupervised learning using the shadow kernel.
The simulated domain shifts include Hamiltonian-parameter mismatch, as well as changes in system size, partition structure, and state-generation procedure, together with imperfect state preparation and hardware-type noise.

For the quantum phase classification benchmarks, UDA attains high target performance even when the target states have nonuniform and sometimes visibly reduced overlap with the ground-state subspace.
In particular, it already achieves strong accuracy at $T_{\tgt}=10^3$ and improves further at $T_{\tgt}=10^4$, with the resulting phase diagrams recovering the overall cluster and ANNNI phase structure over most of the target region, including the gapless floating phase in ANNNI. Although the setups are not directly comparable, these target-domain accuracies are already comparable to recent clean supervised QCNN results \cite{Bermejo2024QCNNSimulable}.
For the toric-code and color-code benchmarks, UDA remains effective even at 98 and 91 qubits with only 500 shots per data point.
Most of the remaining classification errors in the cluster and ANNNI models are concentrated near phase boundaries and in the most difficult parts of the target domain.
Improving robustness in such boundary regions is therefore an important direction for future investigation.

The entanglement classification benchmarks further show that the benefit of UDA can be especially pronounced in practically difficult transfer settings. For multipartite separable versus entangled states classification, UDA markedly improves the target performance, most notably in the severe partition-mismatched transfer $(16,6,\mathrm{fix})$-Clean$\rightarrow(16,3,\mathrm{random})$-Noisy, where the source-only baseline is particularly poor and UDA raises the macro-F1 to around $0.95$ in Table~\ref{table:entangle}. Again, these results are obtained already with only 500 shots per data point. In this task, UDA also clearly outperforms target-only unsupervised clustering with the shadow kernel. By contrast, the GHZ/W benchmarks show a different pattern. UDA is beneficial in the harder size-shifted setting, but target-only unsupervised clustering with the shadow kernel performs even better, and in the $(16,\mathrm{pauli})$-Clean$\rightarrow(16,\mathrm{slocc})$-Noisy setting the source-only model already transfers strongly. This suggests that, for some tasks, exploiting target-side class structure can be more important than domain alignment alone.

At the same time, the results show that the choice of label-free model-selection metric is also important.
In the spin-chain benchmarks, especially for ANNNI, EnsV is more reliable, whereas in the multipartite Sep/Ent task InfoMax performs better and yields the strongest results. Taken together, these results indicate that UDA can remain highly effective even under substantial imperfections, while its final performance depends materially on how the adapted model is selected without target labels.

Several directions deserve further study.
On the methodological side, it will be important to combine domain adaptation with more advanced methods and stronger label-free model-selection strategies to further improve performance and stability \cite{yue2023make,Qu2024ConnectL}.
On the application side, exploring more realistic quantum many-body tasks and larger-scale benchmarks represents a natural next step, including systems for which classical simulation becomes infeasible and quantum hardware is required to generate the data.
Finally, validation on near-term quantum devices will be an important step toward establishing domain adaptation as a practical tool for learning from imperfect quantum data, especially in view of recent progress on classical learning from quantum experimental data in many-body settings \cite{Cho2024QuantumExperimentalData}.

\section*{Acknowledgement}
This work was supported by MEXT Quantum Leap Flagship Program Grants No. JPMXS0118067285 and No. JPMXS0120319794.

\setcounter{equation}{0}
\setcounter{section}{0}

\appendix

\section{Neural-network architectures}
\label{app:network-architectures}

For reproducibility, we list the PyTorch implementation of the CNN feature extractor, classifier, and domain discriminator used in the numerics.
In the actual implementation, each batch normalization layer is replaced by a domain-specific batch normalization layer.

\begin{lstlisting}[caption={PyTorch implementation of the neural-network modules used in the numerics.},label={lst:network-architectures}]
import torch.nn as nn
import torch.nn.functional as F

class FeatureExtractor(nn.Module):
    def __init__(self, input_shape, **kwargs):
        super().__init__()
        # --- CNN -----
        T, C = input_shape
        C_out = 2*(C+1)
        self.conv1 = nn.Conv1d(C, C_out, 3, padding=1)
        self.bn1   = nn.BatchNorm1d(C_out)
        C_in = C_out
        C_out = 2 * C_in
        self.conv2 = nn.Conv1d(C_in, C_out, 3, padding=1)
        self.bn2   = nn.BatchNorm1d(C_out)
        self.pool1 = nn.MaxPool1d(2)
        C_in = C_out
        C_out = 2 * C_in
        self.conv3 = nn.Conv1d(C_in, C_out, 3, padding=1)
        self.bn3   = nn.BatchNorm1d(C_out)
        C_in = C_out
        self.conv4 = nn.Conv1d(C_in, C_out, 3, padding=1)
        self.bn4   = nn.BatchNorm1d(C_out)
        self.pool2 = nn.MaxPool1d(2)
        C_out = 2 * C_in
        self.conv5 = nn.Conv1d(C_in, C_out, 3, padding=1)
        self.bn5   = nn.BatchNorm1d(C_out)
        self.pool3 = nn.MaxPool1d(2)

        length = input_shape[0] // 8

        self.flatten = nn.Flatten()
        feature_dim = C_out // 4
        self.fc1 = nn.Linear(C_out * length, feature_dim)
        self.out_dim = feature_dim

    def forward(self, x):          # x: [B, T, 15]
        x = x.transpose(1, 2)      # -> [B, 15, T]
        x = F.relu(self.bn1(self.conv1(x)))
        x = F.relu(self.bn2(self.conv2(x))); x = self.pool1(x)
        x = F.relu(self.bn3(self.conv3(x)))
        x = F.relu(self.bn4(self.conv4(x))); x = self.pool2(x)
        x = F.relu(self.bn5(self.conv5(x))); x = self.pool3(x)
        x = self.flatten(x)
        return F.relu(self.fc1(x))  # [B, 64]

class Classifier(nn.Module):
    def __init__(self, num_classes: int, input_dim: int):
        super().__init__()
        self.fc = nn.Linear(input_dim, num_classes)

    def forward(self, x):
        return self.fc(x)

class DomainDiscriminator(nn.Module):
    def __init__(self, input_dim):
        super().__init__()
        self.fc = nn.Linear(input_dim, 2)

    def set_fc(self, input_dim):
        self.fc = nn.Linear(input_dim, 2)

    def forward(self, x):
        return self.fc(x)
\end{lstlisting}

\section{Construction of random and QETU target states}
\label{app:target-state-construction}

This appendix gives the detailed construction of the target states used for the algorithmic imperfection model.
Let
\[
\{(E_m(\bm{x}),\ket{\phi_m(\bm{x})})\}_m
\]
denote the computed low-energy eigenpairs at parameter $\bm{x}$, ordered by increasing energy.

Raw target states are constructed first.
We randomly draw the overlap parameter $f$ uniformly from $[0.2, 0.4]$,
sample $\ket{v_g}$ uniformly from the unit sphere of $\mathcal{G}(\bm{x})$, and sample $\ket{v_e}$ uniformly from the unit sphere of the orthogonal complement of $\mathcal{G}(\bm{x})$ within the span of the computed eigenvectors.
We then set
\[
\ket{\psi_{\mathrm{raw}}(\bm{x})}
=
\sqrt{f}\,\ket{v_g}
+
\sqrt{1-f}\,\ket{v_e}.
\]
Equivalently, in the computed eigenbasis,
\[
\ket{\psi_{\mathrm{raw}}(\bm{x})}
=
\sum_m a_m^{(0)}(\bm{x}) \ket{\phi_m(\bm{x})},
\]
where the coefficients satisfy
\[
\sum_{m \in I_{\mathcal{G}}(\bm{x})}
|a_m^{(0)}(\bm{x})|^2
=
f,
\]
and $I_{\mathcal{G}}(\bm{x})$ is the set of indices corresponding to $\mathcal{G}(\bm{x})$.

The second construction is motivated by QETU~\cite{Dong2022QETU}.
In QETU, quantum signal processing of $e^{-i \tilde{H}}$ is used to implement an even polynomial approximation to a step-function spectral filter in the variable $\cos(\widetilde H/2)$, where $\widetilde H$ is a rescaled Hamiltonian whose spectrum lies in $[\eta,\pi-\eta]$.
We take $\eta=0.05$.
In an ideal circuit implementation, obtaining a high-quality ground-state filter requires a polynomial degree that scales as $O(\Delta^{-1}\log \epsilon^{-1})$ in terms of a lower bound $\Delta$ on the spectral gap and target accuracy $\epsilon$, and the same order of queries to the Hamiltonian-simulation primitive.
Moreover, since the filter is implemented nonunitarily, successful preparation is conditioned on the flag-qubit outcome.
In the present numerical experiments, however, we do not simulate this circuit-level postselection process or the implementation error of Hamiltonian simulation.
Instead, we directly apply the induced spectral filter in the computed eigenbasis, assuming that the error of the Hamiltonian simulation is negligible.

To define the filter, we first rescale the Hamiltonian as
\[
\widetilde H(\bm{x})
=
c_1(\bm{x}) H(\bm{x}) + c_2(\bm{x}) I,
\]
with
\begin{equation}
\begin{aligned}
c_1(\bm{x})
&=
\frac{\pi - 2\eta}
{E_{\max}^{\mathrm{est}}(\bm{x}) - E_0^{\mathrm{lb}}(\bm{x})},
\\
c_2(\bm{x})
&=
\eta - c_1(\bm{x}) E_0^{\mathrm{lb}}(\bm{x}),
\end{aligned}
\end{equation}
where $E_0^{\mathrm{lb}}(\bm{x}) = E_0(\bm{x}) - \varepsilon_0$ is a worst-case estimated lower bound of the minimum eigenvalue $E_0(\bm{x})$ of $H(\bm{x})$ when we have its estimate with precision $\varepsilon_0/2$.
We use $\varepsilon_0=0.01$.
Here $E_{\max}^{\mathrm{est}}(\bm{x})$ is taken to be a trivial upper bound on $\|H(\bm{x})\|$.
For the cluster model, we use
\[
E_{\max}^{\mathrm{est}}(\bm{x})
=
n|J| + (n-1)|h_1| + (n-2)|h_2|,
\]
and for the ANNNI model, we use
\[
E_{\max}^{\mathrm{est}}(\bm{x})
=
(n-1)\bigl(|J|+|h_1|\bigr) + n|h_2|.
\]

Next, we specify the transition window of the filter using a fixed gap estimate $\Delta_{\mathrm{est}}$.
We use $\Delta_{\mathrm{est}}=1$ for the cluster model and $\Delta_{\mathrm{est}}=0.01$ for the ANNNI model.
We then define
\begin{equation}
\begin{aligned}
\lambda_0^{\mathrm{est}}(\bm{x})
&=
c_1(\bm{x}) E_0^{\mathrm{ub}}(\bm{x}) + c_2(\bm{x}),
\\
\lambda_1^{\mathrm{est}}(\bm{x})
&=
\lambda_0^{\mathrm{est}}(\bm{x}) + c_1(\bm{x}) \Delta_{\mathrm{est}},
\end{aligned}
\end{equation}
where $E_0^{\mathrm{ub}}(\bm{x})
=
E_0(\bm{x}) + \varepsilon_0$.

The filter itself is taken to be an even Chebyshev polynomial of degree $k_{\rm{deg}}=40$,
\[
P_{40}(z) = \sum_{k=0}^{20} \alpha_k T_{2k}(z),
\]
whose coefficients are obtained by a discrete minimax design on Chebyshev grids.
The design intervals are determined by
\[
z_{\max} = \cos\frac{\eta}{2},
\qquad
z_{\min} = \cos\frac{\pi-\eta}{2},
\]
and
\begin{equation}
\begin{aligned}
z_{+}(\bm{x})
&=
\cos\frac{
\widehat\mu(\bm{x})-\widehat\Delta(\bm{x})/2
}{2},
\\
z_{-}(\bm{x})
&=
\cos\frac{
\widehat\mu(\bm{x})+\widehat\Delta(\bm{x})/2
}{2},
\end{aligned}
\end{equation}
where
\begin{equation}
\begin{aligned}
\widehat\Delta(\bm{x})
&=
0.9\, c_1(\bm{x}) \Delta_{\mathrm{est}},
\\
\widehat\mu(\bm{x})
&=
\frac{
\lambda_0^{\mathrm{est}}(\bm{x}) + \lambda_1^{\mathrm{est}}(\bm{x})
}{2}.
\end{aligned}
\end{equation}
The coefficients are chosen so that $P_{40}(z)$ approximates $1$ on $[z_{+}(\bm{x}),z_{\max}]$ and $0$ on $[z_{\min},z_{-}(\bm{x})]$, under the global constraint $|P_{40}(z)| \le 0.999$ on $[z_{\min},z_{\max}]$.

Finally, define
\begin{equation}
b_m(\bm{x})
=
P_{40}\!\left(
\cos\frac{
c_1(\bm{x}) E_m(\bm{x}) + c_2(\bm{x})
}{2}
\right)
a_m^{(0)}(\bm{x}).
\end{equation}
The QETU target state is then
\begin{equation}
\ket{\psi_{\mathrm{QETU}}(\bm{x})}
=
\frac{
\sum_m b_m(\bm{x}) \ket{\phi_m(\bm{x})}
}{
\left\|
\sum_m b_m(\bm{x}) \ket{\phi_m(\bm{x})}
\right\|
}.
\end{equation}
This normalized filtered state is used for generating $R_{\tgt}$-QETU-$T_{\tgt}$ target dataset in the numerical simulation.






\bibliography{papers,addition}

\end{document}